\documentclass[aps,prd,reprint,superscriptaddress,longbibliography,showpacs,floatfix,twocolumn]{revtex4-2} 
\usepackage{graphicx}

\usepackage{times,bm,url,amsmath,upgreek}
\usepackage{amssymb}
\usepackage[colorlinks=true,breaklinks=true,allcolors=blue]{hyperref}
\usepackage[utf8]{inputenc}

\graphicspath{ {./figures/} }
\newcommand{\orcidicon}[1]{\href{https://orcid.org/#1}{\includegraphics[height=\fontcharht\font`\B]{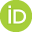}}} 

\usepackage{comment}
\usepackage{subfigure}
\usepackage[scr=boondox]{mathalfa}
\usepackage{xparse}
\usepackage{bbm}
\usepackage{bbold} 

\begin{document}
	
	\title{Pole-skipping in two-dimensional de Sitter spacetime and double-scaled SYK model}
	
	\author{Haiming Yuan \!\orcidicon{0009-0002-0610-7936}}
	\affiliation{Department of Physics, College of Sciences, Shanghai University, 99 Shangda Road, Shanghai 200444, China}
	\affiliation{School of Physics, Henan University of Technology, 100 Lianhua Street, Zhengzhou 450001, China}
 	\author{Xian-Hui Ge \!\orcidicon{0000-0001-6228-0376}}
 	\email[Corresponding author. ]{ gexh@shu.edu.cn}
 	\affiliation{Department of Physics, College of Sciences, Shanghai University, 99 Shangda Road, Shanghai 200444, China}
 	\author{Keun-Young Kim \!\orcidicon{0000-0002-4725-3211}}
 		\affiliation{Department of Physics and Photon Science, Gwangju Institute of Science and Technology, 123 Cheomdan-gwagiro, Gwangju 61005, Korea}
 		\affiliation{Research Center for Photon Science Technology, Gwangju Institute of Science and Technology, 123 Cheomdan-gwagiro, Gwangju 61005, Korea}
	\date{\today}
	\begin{abstract}
	We develop the pole-skipping structure in de Sitter (dS) spacetime and find that their leading frequencies satisfy the relation $\omega_{dS}=i2\pi T_{dS}(1-s)$, where $T_{dS}=1/2\pi L$ and $s$ denotes spin. In the two-dimensional dS spacetime, the pole-skipping points near the cosmic horizon $r=L$ for the scalar field of spin-0 and the fermionic field of spin-$\frac{1}{2}$ correspond one-to-one with those in the classical limit as $\lambda\rightarrow 0$ in double-scaled Sachdev-Ye-Kitaev model when the temperature is infinite (DSSYK$_\infty$). This provides a numerical correspondence between quantum gravity in the static patch of two-dimensional dS spacetime and a one-dimensional quantum system, which we consider as a realization of the DS/dS correspondence.
	\end{abstract}

\maketitle
\section{Introduction}
  The holographic principle, which posits that a theory of quantum gravity in a higher-dimensional spacetime can be encoded by a lower-dimensional boundary theory, is a cornerstone of modern theoretical physics~\cite{Hooft,Susskind5}. One of its most successful realizations is the anti-de Sitter/conformal field theory (AdS/CFT) correspondence~\cite{Maldacena2,gubser,witten1,witten2}, which connects a gravitational theory in an asymptotically AdS spacetime with a CFT living on its boundary. This correspondence has provided deep insights into the interplay between quantum field theories and gravity. 
  
 Attempts have been made to connect quantum gravity in de Sitter spacetime to a lower-dimensional conformal field theory, as the correspondence holds significant promise for understanding quantum gravity in a universe that is expanding, akin to our own. Some papers have proposed the dS/CFT correspondence, which describes that quantum gravity in de Sitter space is dual to CFT living on the past and future boundaries of de Sitter space~\cite{Gibbons,Maldacena4,Strominger,Strominger2,Witten3,Klemm,Harlow,Strominger3,Strominger4,Hikida,Hikida2,Chen}, where the masses of the stable scalar make the conformal weights become complex, hence the boundary CFT is non-unitary. The DS/dS correspondence, as a framework for describing holographic duality in dS gravity, provides us with a new perspective~\cite{Karch,Alishahiha,Geng1,Geng2,Geng3,Geng4}. It describes the duality between quantum gravity in $(d+1)-$dimensional dS spacetime and two CFTs, which are cut off at the energy scale $1/L$ (where $L$ is the radius of dS) and are coupled through dynamical gravity. Both dS$_{d+1}$ and AdS$_{d+1}$ can be represented as foliations of dS$_{d}$ slices, and the quantum gravity theories in dS$_{d+1}$ and AdS$_{d+1}$ have the same IR physics, which is the IR limit of the CFT on the AdS UV boundary. In this paper, by drawing an analogy with the pole-skipping phenomenon in the AdS/CFT correspondence, we obtain the pole-skipping structure in the DS/dS correspondence, enriching the de Sitter holographic duality.

 Pole-skipping is a phenomenon with very interesting properties in the AdS/CFT theory. Generally, the retarded Green's function takes a form
 \begin{equation}
  \label{eq:40}
  G^R(\omega,k)=\frac{b(\omega,k)}{a(\omega,k)}
 \end{equation}
  in the complex momentum space $(\omega,k)$. At a special point $(\omega_\star, k_\star)$ both $a$ and $b$ satisfy $a(\omega_\star, k_\star)=b(\omega_\star, k_\star)=0$, and the retarded Green's function cannot be uniquely defined~\cite{Grozdanov1,Blake1,Blake2,Grozdanov2,Das}. Its value will be determined by how it approaches this special point, that is, it depends on the slope $\delta k/\delta \omega$.
 \begin{equation}
  \label{eq:41}
  G^R=\frac{(\partial_\omega b)_\star +\frac{\delta k}{\delta \omega}(\partial_k b)_\star+\dots}{(\partial_\omega a)_\star +\frac{\delta k}{\delta \omega}(\partial_k a)_\star+\dots}.
 \end{equation}
  So if we find the intersections of zeros and poles in the retarded Green's functions, we can obtain those special points, which we refer to them as pole-skipping points. For the theory of the AdS/CFT correspondence, we can use another method to obtain the pole-skipping points from the bulk field equation \cite{Makoto1,Makoto2,BlakeDavison,Abbasi1,Abbasi2,Abbasi3,Choi,Karunava,Ahn1,Mahdi}. The absence of a unique incoming mode near the horizon corresponds to the non-uniqueness of the Green's function on the boundary. For the static black holes in AdS spacetime, the leading pole-skipping frequency $\omega$ is known as $\omega_{AdS}=i2\pi T_{AdS}(s-1)$~\cite{Makoto3,N1,Yuan1,N2,Yuan2,Diandian,Jeong,Ning,Ahn2},
  where $i$ is the imaginary unit, and $s$ denotes the spin of the operator. The frequencies of the pole-skipping points in dS spacetime are different from those in AdS spacetime. The frequencies of the pole-skipping points of the scalar perturbation near the horizon of a four-dimensional asymptotically dS black hole are given as $\omega=i2\pi T n$ ($n$ is a positive integer)~\cite{Grozdanov3}, which are opposite to the pole-skipping frequencies of the scalar perturbation in AdS spacetime.
  
  In this work, we calculate the pole-skipping positions of the scalar field and the fermionic field in dS spacetime and obtain the leading frequencies $\omega_{dS}=i2\pi T_{dS}(1-s)$, where $T_{dS}$ denotes $T_{dS}=1/2\pi L$, and $s$ is the spin. The temperature in dS space is given as $T_{dS}=1/2\pi L$, where $L$ is the dS radius~\cite{Gibbons,Strominger2}. In dS$_2$ spacetime, only the scalar field with spin-0 and fermionic field with spin-$\frac{1}{2}$ exist, so the leading frequencies of the pole-skipping point are only located in the upper half-plane. For this special case in dS$_2$ gravity, we aim to identify a dual field theory where the imaginary frequencies of pole-skipping points also reside in the upper half-plane.
  
  The Sachdev-Ye-Kitaev (SYK) model is a strongly interacting quantum system at low energy which can be solvable at a large $N$ limit with an emergent conformal symmetry~\cite{Sachdev,Kitaev}. Both the Jackiw-Teitelboim (JT) gravity with nearly AdS$_2$ geometry and the Majorana SYK model could derive the Schwarzian action by spontaneously breaking their conformal symmetry down to $SL(2,\mathbb{R})$ due to reparametrization~\cite{Jackiw,Teitelboim,Sarosi,Stanford,Polchinski,Roberts,Tarnopolsky,Cai1,Cai2,Cao,Cai3,Cao2}. Therefore, AdS JT gravity has been considered as the gravitational dual of the Majorana SYK model. In Ref.~\cite{Yuan3}, we show that the pole-skipping points in the Majorana SYK model match those in AdS JT gravity. Moreover, the pole-skipping points in the charged AdS JT gravity and in the complex SYK model remain the same. This demonstrates the dualities JT/Majorana SYK and charged of JT/complex SYK in terms of the pole-skipping. 
  
  The double-scaled SYK (DSSYK) model is the limit of the SYK model under conditions $N,\,p\rightarrow \infty$ and $\lambda\equiv\frac{2p^2}{N}=\text{fixed}$~\cite{Cotler,Berkooz,Berkooz2,Khramtsov}. One can compute the exact 4-point function for the DSSYK model across all energy scales using combinatorial methods and chord diagrams. It applies these results to correct the maximal Lyapunov exponent at low-temperatures and introduces diagrammatic rules for non-perturbative correlation functions. The double scale limit simplifies some calculations and makes it easier to analyze and study the dynamical and spectral properties of correlation functions. It is an important step in understanding the solvable model of holographic duality. 
  
  In the classical limit and low-temperature limit, the behavior of DSSYK model matches that of the Schwarzian theory~\cite{Berkooz,Lin,Berkooz3}, which describes the dynamics of Jackiw-Teitelboim (JT) gravity in two-dimensional AdS space. However, interestingly, the DSSYK model exhibits different behavior in the high-temperature limit. Recently, a pioneering work by Susskind and collaborators established a connection between a static patch of JT gravity in dS spacetime and the double-scaled Sachdev-Ye-Kitaev model at infinite-temperature (DSSYK$_\infty$)~\cite{Saad,Maldacena3,Cotler2,Susskind6,Cotler3}. JT gravity in nearly 2D dS spacetime can be described using a double-scaled random matrix theory, which shares properties similar to those of the DSSYK model. In the double-scaled limit, the spectral form factors of the SYK model exhibit the general ramp-plateau behavior of random matrix theory, suggesting a potential holographic duality between a static patch of JT dS$_2$ gravity and the DSSYK$_\infty$ model~\cite{Susskind1,Susskind2,Susskind3,Rahman,Susskind4,Narovlansky,Verlinde}. In~\cite{Jiuci1,Jiuci2}, they presented a complete solution to the Brownian version of DSSYK and explored its potential relation to static patch physics in de Sitter space. In particular, they derived an analytic result for the OTOC, which exhibits hyper-fast scrambling in the semi-classical limit. They demonstrated that the double-scaled chord algebra forms a Type ${\rm II}_1$ von Neumann algebra, which explains why the empty chord state corresponds to an infinite-temperature state. Refs.~\cite{Geng2,Geng3,Geng4} have understood the fact that the de Sitter static patch is in a maximally entangled state, and this fact supports the recent discussion of Type ${\rm II}_1$ algebra in de Sitter space.
  
  We aim to explore the numerical connection between two-dimensional dS spacetime and the DSSYK$_\infty$ model in this paper. We find that, in the classical limit as $\lambda\rightarrow 0$, the pole-skipping points of the Green's function in the DSSYK model correspond one-to-one with these special points at the cosmic horizon which is the boundary of the static patch in dS$_2$ spacetime. The duality between AdS$_2$ and the low-temperature SYK model could lead to a one-to-one correspondence in their pole-skipping structures~\cite{Yuan3}. Therefore, we believe that it has a dual relationship with dS$_2$ spacetime through the pole-skipping phenomenon. Our findings support the conjecture of a holographic duality between the static patch of JT dS$_2$ gravity and the DSSYK$_\infty$ model. 
  
  In Sec.~\ref{sec:1} and Sec.~\ref{sec:2}, we calculate the pole-skipping points for the scalar field of spin-0 and the fermionic field of spin-$\frac{1}{2}$ at the dS$_2$ horizon by using the methods of near-horizon analysis and computing bulk retarded Green's function separately. In Sec.~\ref{sec:3}, we calculated the pole-skipping points of the DSSYK model in the high-temperature limit, and these results correspond to those in two-dimensional dS spacetime. We summarize and discuss in Sec.~\ref{sec:Conclusion}.
	
  \section{Pole-skipping in dS$_2$ spacetime: near-horizon analysis} \label{sec:1}
   \subsection{Scalar field}
   \label{sec:11}
  The two-dimensional JT gravity with positive cosmological constant is a dimensional reduction of three-dimensional Einstein gravity with a positive cosmological constant, and the metric of dS$_2$ with a dilaton field $\Phi$ equal to $r$ is given as~\cite{Susskind3,Rahman,Susskind4}
  \begin{equation}
    \begin{aligned}
   \label{eq:1}
   ds^2=-f(r)dt^2+\frac{1}{f(r)}dr^2,\quad\Phi=r,
   	\end{aligned}
  \end{equation}
   where $f(r)=1-r^2/L^2$ and $L$ is the radius of the dS. The cosmic horizon at $r_h=L$, and the global temperature is given by $T_{dS}=\frac{1}{2\pi L}$~\cite{Gibbons,Strominger2}. We consider the scalar field $\Psi(v,r)=e^{-i \omega t}\Psi(r)$ with mass $m$ of which the dynamics are given by the Klein-Gordon equation
    \begin{equation}
   \label{eq:2}
   \frac{1}{\sqrt{-g}}\partial_\mu(\sqrt{-g}g^{\mu\nu}\partial_\nu\Psi)-m^2\Psi=0\,.
   \end{equation}
   We use the incoming Eddington-Finkelstein (EF) coordinate $v=t+r_*$, where $r_*$ is the tortoise coordinate $dr_*=dr/f(r)$. In the incoming EF coordinate,~\eqref{eq:1} becomes
    \begin{equation}
   \label{eq:3}
   ds^2=-f(r)dv^2+2dvdr,
   \end{equation}
   and the Klein-Gordon~\eqref{eq:2} equation can be expanded as
    \begin{equation}
   \label{eq:4}
   f(r)\Psi''(r)+(f'(r)-2i\omega)\Psi'(r)-m^2\Psi(r)=0.
    \end{equation}
   Since the blacken factor goes $f(r) = 4\pi T(r_h-r) + \mathcal{O}\left(r_h-r\right)^2$ near the horizon, one can check that the Klein-Gordon equation~\eqref{eq:4} has a regular singularity at $r=r_h$. From now on we will use scaled frequency $\mathfrak{w}=\frac{\omega}{2\pi T}$, and mass $\mathfrak{m}=\frac{m}{2\pi T}$. The singularity can be seen if we approximate~\eqref{eq:4} to
    \begin{equation}
   \label{eq:5}
   \Psi''(r)-\frac{1+i\mathfrak{w}}{r_h-r}\Psi'(r)-\frac{\pi T\mathfrak{m}^2}{r_h-r}\Psi(r)=0,
   \end{equation}
   near horizon. According to conventional differential equation techniques, one can solve \eqref{eq:5} by imposing series solution ansatz as
    \begin{equation}
   \label{eq:6}
   \Psi(r)=(r-r_h)^\lambda\sum^\infty_{p=0}\Psi_p(r-r_h)^p.
   \end{equation}
   At the lowest order, we can obtain the indicial equation $\lambda(\lambda+i\mathfrak{w})=0$. The two roots are
   \begin{equation}
   \label{eq:7}
   \lambda_1=0,\quad \lambda_2=-i\mathfrak{w}\,.
   \end{equation}
   Generally speaking, a solution with exponent $\lambda_1$ near the horizon is regular. Therefore, if $\lambda_2$ is not an integer, we can obtain a unique ``incoming" solution with exponent $\lambda_1$. However, if $\lambda_2$ is an integer, there may be an additional regular solution, which is a signal of a pole-skipping near the horizon. For example, let us consider $i\mathfrak{w}=-1$. According to the standard technique of the differential equation, the assumption for the following solution is more appropriate than~\eqref{eq:6}
    \begin{equation}
   \label{eq:8}
   \Psi(r)=\sum^\infty_{p=0}\Psi_{1,p}(r-r_h)^p + (r-r_h)\log(r-r_h)\sum^\infty_{q=0}\Psi_{2,q}(r-r_h)^q\,,
   \end{equation}
   where we include $\log$ term. After substituting it into the Klein-Gordon eqaution~\eqref{eq:4}, up to $\mathcal{O}(r-r_h)^0$ the equation of motion becomes
   \begin{equation}
   	\begin{aligned} 	
   \label{eq:9}
   &(1+i\mathfrak{w})\left(\Psi_{1,1}+ \Psi_{2,0}\log(r-r_h)\right)+\big(\pi T \mathfrak{m}^2\Psi_{1,0}\\
   &+(2+i \mathfrak{w})\Psi_{2,0}\big)=0\,.
\end{aligned}
   \end{equation}
   The first term vanishes at $i\mathfrak{w}=-1$. If we focus on $\mathfrak{m}=0$, we obtain $\Psi_{2,0}=0$, and the series solution takes two independent regular solutions as follows
   \begin{equation}
   \label{eq:10}
   \Psi(r)=\Psi_{1,0} + (r-r_h)^1\sum^\infty_{p=0}\tilde{\Psi}_{2,p}(r-r_h)^p\,,
   \end{equation}
   where two independent coefficients are $\Psi_{1,0}$ and $\tilde{\Psi}_{2,0}$. Thus the first pole-skipping point is
   \begin{equation}
   \label{eq:11}
   \mathfrak{w}_{\star}=-i,\quad \mathfrak{m}^2_{\star}=0\,.
   \end{equation}
   In general, at the pole-skipping points, the series solution takes the following form
   \begin{equation}
   \label{eq:12}
   \Psi(r)=\sum^{n-1}_{p=0}\Psi_{1,p}(r-r_h)^p + (r-r_h)^n\sum^\infty_{q=0}\tilde{\Psi}_{2,q}(r-r_h)^q\,,
   \end{equation}
   where $\Psi_{1,p}$ and $\tilde{\Psi}_{2,q}$ are independent coefficients. There is a systematic procedure to calculate pole-skipping points~\cite{BlakeDavison}. Firstly, we expand $\Psi(r)$ with a Taylor series
   \begin{equation}
   \label{eq:13}
   \Psi(r)=\sum^\infty_{p=0}\Psi_p(r-r_h)^p=\Psi_0+\Psi_1(r-r_h)+\Psi_2(r-r_h)^2+\dots.
   \end{equation}
   We substitute \eqref{eq:13} into \eqref{eq:4} and expand the equation of motion in powers of $(r-r_h)$. Then, a series of the perturbed equation in the order of $(r-r_h)$ can be denoted as
   \begin{equation}
   \label{eq:14}
   S=\sum^\infty_{p=0}S_p(r-r_h)^p=S_0+S_1(r-r_h)+S_2(r-r_h)^2+\cdots\,.
   \end{equation}
   We write down the first few equations $S_p=0$ in the expansion of~\eqref{eq:14}:
   \begin{equation}
   	 	\begin{aligned} 
   		0=&M_{11}(\omega,m)\Psi_0+(2\pi T+i\omega)\Psi_1,\\
   		0=&M_{21}(\omega,m)\Psi_0+M_{22}(\omega,m)\Psi_1+(4\pi T+i\omega)\Psi_2,\\
   		0=&M_{31}(\omega,m)\Psi_0+M_{32}(\omega,m)\Psi_1+M_{33}(\omega,m)\Psi_2\\
   		&+(6\pi T+i\omega)\Psi_3\,.
   	\end{aligned}
   \end{equation}
   To find an incoming solution, we should solve a set of linear equations of the form
   \begin{equation}
   		\begin{aligned}
   \label{eq:15}
   &\mathcal{M}(\omega,m)\cdot \Psi\equiv\\
   &\left(\begin{array}{ccccc}
   	M_{11} & (2\pi T+i\omega) & 0    & 0  &\dots\\
   	M_{21} & M_{22}& (4\pi T+i\omega)& 0   &\dots\\
   	M_{31} & M_{32}&  M_{33} &(6\pi T+i\omega) &\dots\\
   	\dots   &  \dots&  \dots  &\dots   &\dots\\
   \end{array}\right)\\
   &\times\left(\begin{array}{ccccc}
   	\Psi_0\\
   	\Psi_1\\
   	\Psi_3 \\
   	\dots \\
   \end{array}\right)=0\,.
   	\end{aligned}
   \end{equation}
   The $n$-th pole-skipping points $(\omega_{n}, m_{n})$ can be calculated by solving
   \begin{equation}
   \label{eq:16}
   \omega_{\star n}=i2\pi Tn,\qquad {\rm det}\mathcal{M}^{(n)}(\omega_\star,m_\star)=0\,,
   \end{equation}
   where the matrix $\mathcal{M}^{(n)}$ is the $(n\times n)$ square matrix whose elements are taken from $M_{11}$ to $M_{nn}$ in~\eqref{eq:15}. (The first few elements of this matrix have been shown in Appendix~\ref{sec:Details1}.)
   The following are the resulting pole-skipping points:
   \begin{equation}
   \label{eq:17}
   \begin{split}
   	\mathfrak{w}&=i, \quad \mathfrak{m}^2=0\,;\\
   	\mathfrak{w}&=2i, \quad \mathfrak{m}^2=0,-2\,;\\
   	\mathfrak{w}&=3i, \quad \mathfrak{m}^2=0,-2,-6\,;\\
   	\mathfrak{w}&=4i, \quad \mathfrak{m}^2=0,-2,-6,-12\,;\\
   	\mathfrak{w}&=5i, \quad \mathfrak{m}^2=0,-2,-6,-12,-20\,;\\
   	&\qquad \qquad \vdots
   \end{split}
   \end{equation}
   These special points have physical meaning only for the points with mass equal to $0$. Although the other points with negative mass squared are non-physical, their correspondence with the pole-skipping in the later DSSYK model is quite interesting. The operator dimension of scalar field in dS$_2$ spacetime is $\Delta_{1}=\frac{1}{2}+\frac{1}{2}\sqrt{1-4m^2L^2}$~\cite{Strominger,Strominger2,Klemm,Harlow}. Pole-skipping points~\eqref{eq:17} can be rewritten in terms of operator dimensions:
   \begin{equation}
   \label{eq:18}
   \begin{split}
   	\mathfrak{w}&=i, \quad \Delta_{1}=1\,;\\
   	\mathfrak{w}&=2i, \quad \Delta_{1}=1,2\,;\\
   	\mathfrak{w}&=3i, \quad \Delta_{1}=1,2,3\,;\\
   	\mathfrak{w}&=4i, \quad \Delta_{1}=1,2,3,4\,;\\
   	\mathfrak{w}&=5i, \quad \Delta_{1}=1,2,3,4,5\,;\\
   	&\qquad \qquad \vdots
   \end{split}
   \end{equation}
   This result agrees with the formula $\omega_{dS}=i2\pi T_{dS}(1-s)$ for $s=0$. We plot~\eqref{eq:18} in Figure~\ref{fig:Figure1}.
   
    In dS spacetime, the relationship between a scalar field and the conformal dimension of the dual CFT operator is governed by the equation $\Delta(\Delta-d)=-m^2L^2$. When $m^2L^2>1$, the conformal dimension $\Delta$ becomes complex, which implies that the corresponding CFT violates unitarity if there exist stable scalars with masses above this bound. However, for the pole-skipping points we obtained, the squared masses are negative, ensuring that the conformal dimension remains real and thus preserving unitarity in the CFT. This also provides a reasonable explanation for the correspondence with the DSSYK model that satisfies the unitarity in the following steps.

   \subsection{Fermionic field}
   \label{sec:12}
   \quad As a way to confirm the validity of our holographic computational method, we consider the Dirac field in the same background~\eqref{eq:3}. The Dirac equation is given as
   \begin{equation}
   \label{eq:20}
   (\Gamma^MD_M-m)\psi_{\pm}=0.
   \end{equation}
   The capital letter $M$ denotes the indices of bulk spacetime coordinates and small letters $a,b$ denote tangent space indices. The covariant derivative of bulk spacetime acting on fermions is defined by $D_M=\partial_M+\frac{1}{4}(\omega_{ab})_M\Gamma^{ab}$, where $\Gamma_{ab}\equiv\frac{1}{2}[\Gamma_a,\Gamma_b]$. $\Gamma_a$ are Gamma matrices which satisfy Grassman algebra $\{\Gamma^a,\Gamma^b\}=2\eta^{ab}$ \cite{N1,Cai}. The spinors are two dimensional $\psi_{\pm}(r,t)=e^{-i\omega t}\left(
   \begin{array}{c}
   	\psi_{+}(r)\\
   	\psi_{-}(r)\\
   \end{array}
   \right)$ and we use the following gamma matrices representation~\cite{Wilczek}
   \begin{equation}
   \label{eq:21}
   \Gamma^{\underline{v}}=i\sigma^2,\quad \Gamma^{\underline{r}}=\sigma^3.
   \end{equation}
   We choose the orthonormal frame to be
   \begin{equation}
   \label{eq:22}
   E^{\underline{v}}=\frac{1+f(r)}{2}dv-dr,\quad E^{\underline{r}}=\frac{1-f(r)}{2}dv+dr,
   \end{equation}
   for which
   \begin{equation}
   \label{eq:23}
   ds^2=\eta_{ab}E^aE^b,\quad \eta_{ab}={\rm diag}(-1,1).
   \end{equation}
   The spin connections for this frame are given by
   \begin{equation}
   \label{eq:24}
   \omega_{\underline{rr}}=0,\quad \omega_{\underline{vr}}=-\frac{f'(r)}{2}.
   \end{equation}
   Using these spin connections \eqref{eq:24} and the EF coordinate
   \begin{equation}
   \label{eq:25}
   ds^2=-f(r)dv^2+2dvdr,
   \end{equation}
   one can calculate the Dirac equation to be
   \begin{equation}
   \label{eq:26}
   \left\{
   \begin{aligned}
   	&\big(f'(r)-4(m+i\omega)\big)\psi_+(r)+\big(f'(r)-4i\omega\big)\psi_-(r)\\
   	&+2\big(f(r)-1\big)\psi'_-(r)+2\big(f(r)+1\big)\psi'_+(r)=0,\\
   	&\big(4i\omega-f'(r)\big)\psi_+(r)-\big(4m+f'(r)-4i\omega\big)\psi_-(r)\\
   	&-2\big(f(r)+1\big)\psi'_-(r)+2\big(1-f(r)\big)\psi'_+(r)=0.
   \end{aligned}
   \right.
   \end{equation}
   We combine the two equations of \eqref{eq:26} and expand them near the horizon $r_h$. The first-order equation near the horizon is
   \begin{equation}
   \label{eq:27}
   1st:\quad(2\pi T+2i\omega+m)\psi_++(2\pi T+2i\omega-m)\psi_-=0.
   \end{equation}
   We take the value of coefficients $(2\pi T+2i\omega+m)$ and $(2\pi T+2i\omega-m)$ to be 0 and thus there are two independent free parameters $\psi_+$ and $\psi_-$ to this equation. The first-order pole-skipping point is obtained as
   \begin{equation}
   \label{eq:28}
   \mathfrak{w}=\frac{i}{2}, \quad \mathfrak{m}=0, \quad \Delta_{2}=\frac{1}{2},
   \end{equation}
   where the operator dimension of Dirac field in dS$_2$ is $\Delta_{2}=\frac{1}{2}+ im$ which is positive like scalar operator dimension~\cite{Maldacena,Pethybridge}. We expand the Dirac equation in higher order
   \begin{equation}
   	 \begin{aligned}
   \label{eq:29}
   &2nd:\\
   &\bigg(\frac{m^2}{4\pi T}-\frac{m(m-\pi T)}{m-2\pi T-2i\omega}\bigg)\psi^0_{+}+\frac{1}{2}\bigg(3+\frac{i\omega}{\pi T}\bigg)\psi^1_{+}=0;\\
   &3th:\\
   &\frac{(14m\pi T+2im\omega-36\pi^2T^2-m^2)}{(3\pi T+i\omega)(5\pi T+i\omega)(2\pi T+2i\omega-m)}\psi^0_{+}\\
   &\times\frac{m(20m^2\pi^2T^2+64\pi^4T^4+m^4)}{32\pi T}+\frac{3}{2}\big(7+\frac{i\omega}{\pi T}\big)\psi^3_{+}=0;\\
   &4th:\\
   &\frac{m(56m^4\pi^2T^2+2304\pi^6T^6+784m^2\pi^4T^4+m^6)}{(3\pi T+i\omega)(5\pi T+i\omega)(7\pi T+i\omega)(2\pi T+2i\omega-m)}\psi^0_{+}\\
   &\times\frac{(m^2+64\pi^2T^2-18m\pi T-2im\omega)}{192\pi T}+2\big(9+\frac{i\omega}{\pi T}\big)\psi^4_{+}\\
   &=0.
    \end{aligned}
   \end{equation}
   For the choice of positive square root $m$ in higher-order equations, the higher-order pole-skipping points are
   \begin{equation}
   \label{eq:30}
   \begin{split}
   	\mathfrak{w}&=\frac{3i}{2}, \quad \mathfrak{m}=0,i, \quad \Delta_{2}=\frac{1}{2},\frac{3}{2}\,;\\
   	\mathfrak{w}&=\frac{5i}{2}, \quad \mathfrak{m}=0,i,2i, \quad \Delta_{2}=\frac{1}{2},\frac{3}{2},\frac{5}{2}\,;\\
   	\mathfrak{w}&=\frac{7i}{2}, \quad \mathfrak{m}=0,i,2i,3i, \quad \Delta_{2}=\frac{1}{2},\frac{3}{2},\frac{5}{2},\frac{7}{2}\,;\\
   	\mathfrak{w}&=\frac{9i}{2}, \quad \mathfrak{m}=0,i,2i,3i,4i, \quad \Delta_{2}=\frac{1}{2},\frac{3}{2},\frac{5}{2},\frac{7}{2},\frac{9}{2}\,;\\
   	&\qquad \qquad \vdots
   \end{split}
   \end{equation}
   This result agrees with the formula $\omega_{dS}=i2\pi T_{dS}(1-s)$ for $s=1/2$. We plot~\eqref{eq:30} in Figure~\ref{fig:Figure2}. At the position of the pole-skipping points, we cannot distinguish between incoming and outgoing waves at the horizon, which reflects the non-uniqueness of the pole-skipping phenomenon. In the next section, we will see the non-uniqueness of the retarded Green's function at these special points. Similar to the case of the scalar field in the previous section, the points in~\eqref{eq:30} will correspond to the pole-skipping in the later DSSYK model. Therefore, we will still list these points that have the mathematical form of the pole-skipping structure.

\section{Pole-skipping in dS$_2$ spacetime: bulk retarded Green's function}
\label{sec:2}
  In this section, we draw a direct analogy to the AdS/CFT correspondence by using a similar prescription to derive the bulk retarded Green's function in dS spacetime. The bulk fields $\phi$ and $\psi$ are dual to operators in the CFT on the boundary of the static patch with dimensions $\Delta_1$ and $\Delta_2$, respectively. Then we could obtain the pole-skipping points from the retarded Green's function. 
\subsection{Scalar field}
 \label{sec:22}
\begin{figure}[tbp]
  	\centering
  	\subfigure[]{
  		\includegraphics[width=4.2cm]{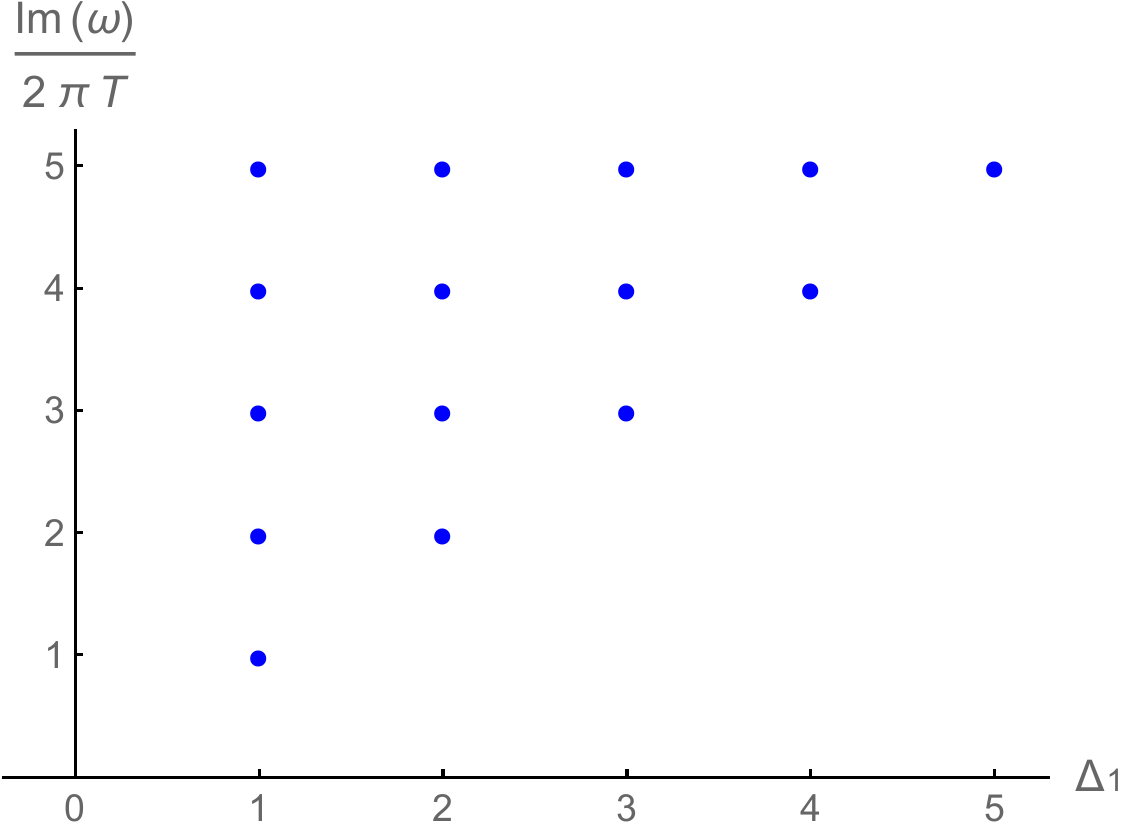}}
  	\subfigure[]{
  		\includegraphics[width=4.2cm]{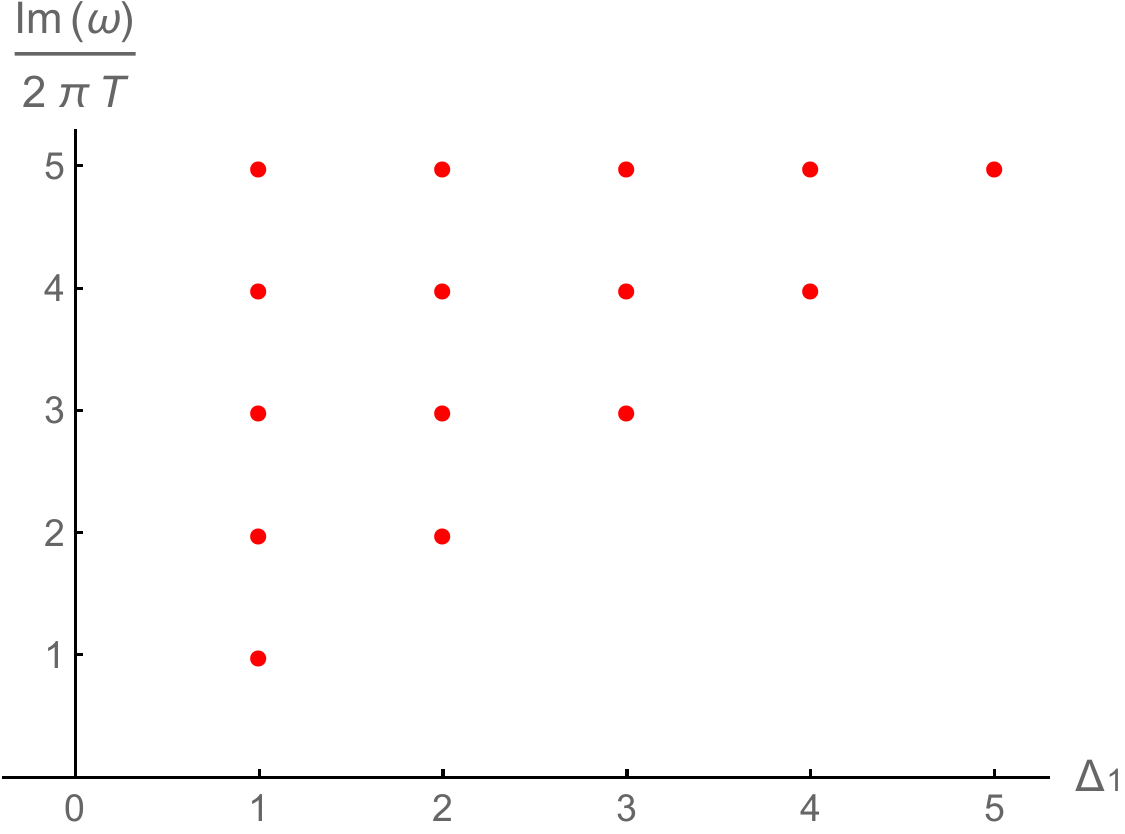}}
  		  	\caption{\label{fig:Figure1} (a) The blue points are the positions of pole-skipping points for the scalar field of the dS$_2$ spacetime by using near horizon analysis;\; (b) The red points are the positions of pole-skipping points for the scalar field of the dS$_2$ spacetime by bulk retarded Green's function.}
 \end{figure}

 We return to the wave equation
  \begin{equation}
  \label{eq:43}
  \frac{1}{\sqrt{-g}}\partial_\mu(\sqrt{-g}g^{\mu\nu}\partial_\nu\phi)-m^2\phi=0\,
  \end{equation}
  in the two-dimensional dS spacetime
  \begin{equation}
  \label{eq:44}
  ds^2=-f(r)dv^2+2dvdr
  \end{equation}
  considered in the previous subsection. The general solution to the wave equation has the form
  \begin{equation}
  \begin{aligned}
  \label{eq:45}
  \phi&=B\phi_{\rm n.}+A\phi_{\rm n.n.}\\
  &\sim  C_1\, P^{(\sqrt{1-4m^2L^2}-1)/2}_{i\omega L}(\frac{r}{L})+C_2\, Q^{(\sqrt{1-4m^2L^2}-1)/2}_{i\omega L}(\frac{r}{L}),
  \end{aligned}
  \end{equation}
  where $P^\mu_\nu(x)$ is the associated Legendre polynomial, and $Q^\mu_\nu(x)$ is the associated Legendre function of the second kind. The functions $P^\mu_\nu(x)$ and $Q^\mu_\nu(x)$ can be expressed in the form of hypergeometric functions~\cite{Whittaker,Silverman}
  	\begin{equation}
  	P^\mu_\nu(x)=\frac{1}{\Gamma(1-\mu)}\bigg(\frac{x+1}{x-1}\bigg)^{\mu/2}\,_2{\bf F}_1\bigg(-\nu,\nu+1;1-\mu;\frac{1-x}{2}\bigg),\nonumber
  	\end{equation}
  	and
  	\begin{equation}
  		\begin{aligned}
  	Q^\mu_\nu(x)=&\frac{(-1)^\mu\sqrt{\pi}\Gamma(\nu+\mu+1)(x^2-1)^{\mu/2}}{2^{\nu+1}\Gamma(\nu+3/2)x^{\nu+\mu+1}}\\
  	&\times\,_2{\bf F}_1\bigg(\frac{\nu+\mu+2}{2},\frac{\nu+\mu+1}{2};\nu+\frac{3}{2};x^{-2}\bigg).
  		\end{aligned}
  	\end{equation}
  We call $\phi_{\rm n.}$ normalizable and $\phi_{\rm n.n.}$ non-normalizable. The specific forms of the two solutions are
  \begin{equation}
  	\begin{aligned}
  \label{eq:46}
  \phi_{\rm n.}&=\frac{\big(1+\frac{r}{L}\big)^{\frac{\sqrt{1-4m^2L^2}-1}{2}}\big(\frac{r^2}{L^2}-1\big)^{\frac{1-\sqrt{1-4m^2L^2}}{4}}}{\Gamma\big(\frac{3-\sqrt{1-4m^2L^2}}{2}\big)}\\
  &\times \,_2{\bf F}_1\bigg(-i\omega L,i\omega L+1;\frac{1}{\Gamma\big(\frac{3-\sqrt{1-4m^2L^2}}{2}\big)};\frac{1}{2}-\frac{r}{2L}\bigg),\\
  \phi_{\rm n.n.}&=\frac{(-1)^{\frac{\sqrt{1-4m^2L^2}-1}{2}}\sqrt{\pi}\Gamma(\frac{\sqrt{1-4m^2L^2}+2i\omega L}{2})}{\Gamma(i\omega L+\frac{3}{2})2^{i\omega L+1}}\\
  &\times \,_2{\bf F}_1\bigg(\frac{1+\sqrt{1-4m^2L^2}+2i\omega L}{4},\\
  &\frac{3+\sqrt{1-4m^2L^2}+2i\omega L}{4};i\omega L+\frac{3}{2};\frac{L^2}{r^2}\bigg)\\
  &\times\big(\frac{r}{L}\big)^{\frac{-1-\sqrt{1-4m^2L^2}-2i\omega L}{2}}\big(\frac{r^2}{L^2}-1\big)^{\frac{\sqrt{1-4m^2L^2}-1}{4}}.
  \end{aligned}
  \end{equation}
  We could expect to treat the non-normalized mode as the source and the normalized mode as the response. Consequently, we can anticipate that the retarded Green's function will be proportional to the ratio $B/A$. We expand the hypergeometric functions in the solutions~\eqref{eq:46} near the horizon, as $r\rightarrow L$, the retarded Green's function becoming (For the hypergeometric function $_2{\bf F}_1\big(a,\,b;\,a+b+\mathcal{N};\,z\big)$ around $z=1$, the details are presented in Appendix~\ref{sec:Details2})
  \begin{equation}
  	\begin{aligned}
  \label{eq:48}
  G^R(\omega,m)\propto\frac{B}{A}\propto&\frac{(-1)^{\frac{\sqrt{1-4m^2L^2}-1}{2}}}{2^{\sqrt{1-4m^2L^2}}}\frac{\Gamma(\frac{3}{2}-\frac{\sqrt{1-4m^2L^2}}{2})}{\Gamma(\frac{3}{2}-\frac{\sqrt{1-4m^2L^2}}{2}+i\omega L)}\\
  &\times\Gamma(\frac{1}{2}+\frac{\sqrt{1-4m^2L^2}}{2}+i\omega L).
   \end{aligned}
  \end{equation}
  Since $\Gamma(-n)$ diverges for non-negative integer $n$, the lines of pole and zero of the Green's function~\eqref{eq:48} are
  \begin{equation}
  \label{eq:49}
  \left\{
  \begin{aligned}
  	\frac{1}{2}+\frac{\sqrt{1-4m^2L^2}}{2}+i\omega L&=0,-1,-2,\cdots \;\text{(lines of pole)}\,,\\
  	\frac{3}{2}-\frac{\sqrt{1-4m^2L^2}}{2}+i\omega L&=0,-1,-2,\cdots\;\text{(lines of zero)}\,.
  \end{aligned}
  \right.
  \end{equation}
  The pole-skipping points for~\eqref{eq:49} are  
  \begin{equation}
  	\label{eq:50}
  	\begin{split}
  		\mathfrak{w}&=i, \quad \Delta_1=1\,;\\
  		\mathfrak{w}&=2i, \quad  \Delta_1=1,2\,;\\
  		\mathfrak{w}&=3i, \quad \Delta_1=1,2,3\,;\\
  		\mathfrak{w}&=4i, \quad \Delta_1=1,2,3,4\,;\\
  		\mathfrak{w}&=5i, \quad \Delta_1=1,2,3,4,5\,;\\
  		&\qquad \qquad \vdots
  	\end{split}
  \end{equation}
  Where $\Delta_1=\frac{1}{2}+\frac{1}{2}\sqrt{1-4m^2L^2}$ is the operator dimension of scalar field mentioned in the Sec~\ref{sec:11}. The results~\eqref{eq:50} agree with the pole-skipping points obtained from near-horizon behavior~\eqref{eq:18}, as shown in Figure~\ref{fig:Figure1}.
  
\subsection{Fermionic field}
 \label{sec:22}
\begin{figure}[tbp]
  	\centering
  	\subfigure[]{
  		\includegraphics[width=4.2cm]{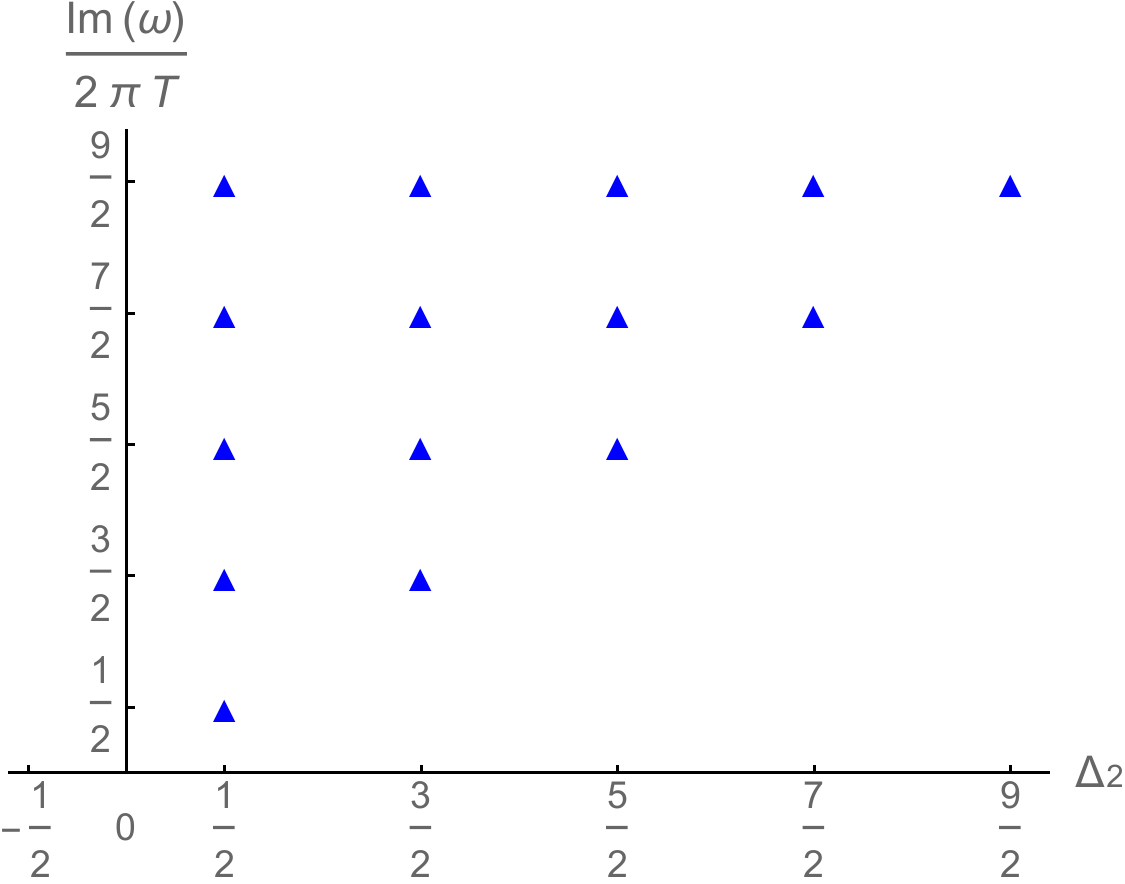}}
  	\subfigure[]{
  		\includegraphics[width=4.2cm]{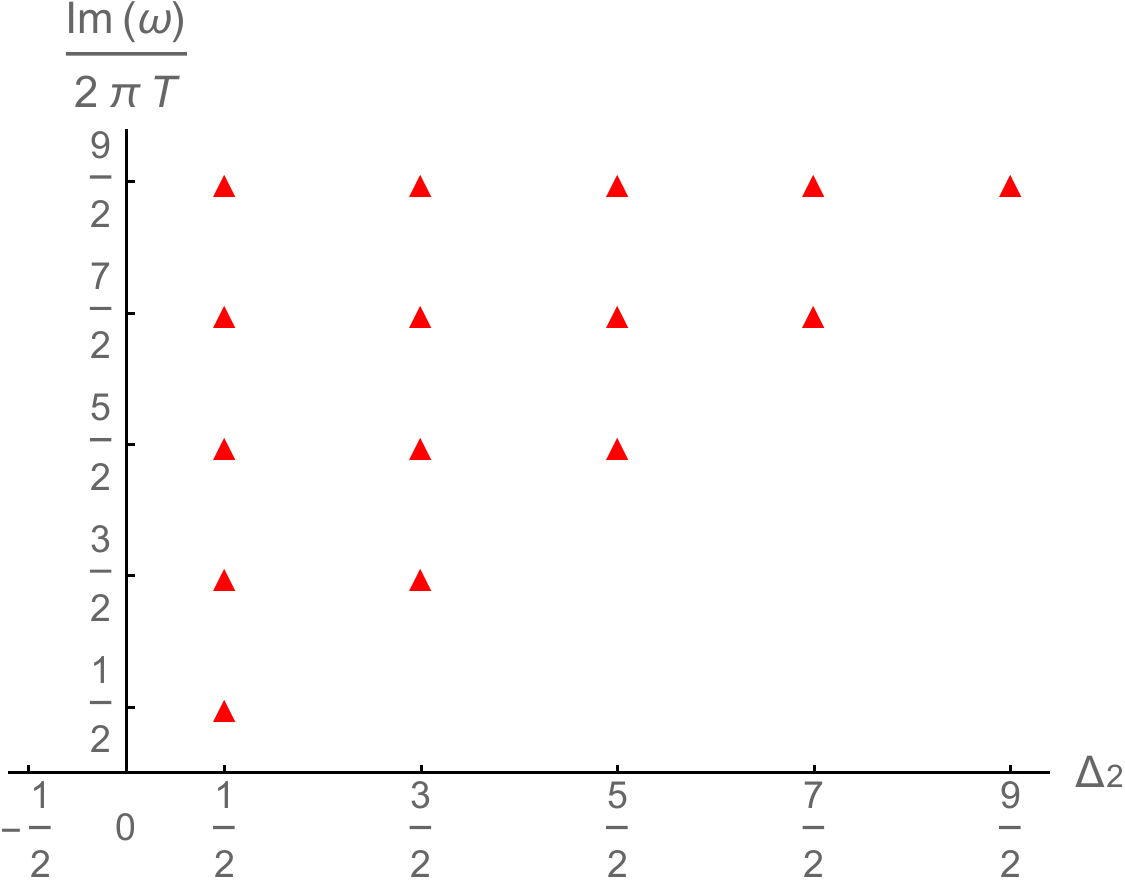}}
  	\caption{\label{fig:Figure2} (a) The blue triangles are the positions of pole-skipping points for the fermionic field of the dS$_2$ spacetime by using near horizon analysis;\; (b) The red triangles are the positions of pole-skipping points for the fermionic field of the dS$_2$ spacetime by bulk retarded Green's function.}
\end{figure}

To solve the fluctuation equation analytically, it is convenient to introduce coordinate $u$, which is defined as $r=L {\rm cos}(u)$. In the $u$ coordinate, metric \eqref{eq:1} takes following form
  \begin{equation}
  \label{eq:51}
  ds^2=-{\rm sin}^2(u)dt^2+L^2du^2,
  \end{equation}
  We substitute the metric~\eqref{eq:51} into the Dirac equation \eqref{eq:23}
  \begin{equation}
  \label{eq:52}
  \left\{
  \begin{aligned}
  	&2\omega L{\rm Csc}(u)\psi_-(u)+({\rm Coth}(u)-2mL)\psi_+(u)\\
  	&+2\psi'_+(u)=0,\\
  	&-2i\omega L{\rm Csc}(u)\psi_+(u)+({\rm Coth}(u)+2mL)\psi_-(u)\\
  	&+2\psi'_-(u)=0.
  \end{aligned}
  \right.
  \end{equation}
  Combining the two equations of \eqref{eq:52}, we obtain
  \begin{equation}
  \label{eq:53}
   \left\{
  \begin{aligned}
  	&\big(4(\omega^2L^2{\rm Csc}^2(u)-m^2L^2-mL{\rm Coth}(u))-2{\rm Csch}^2(u)\\
  	&+{\rm Coth}^2(u)+2{\rm Coth}(u){\rm Cot}(u)\big)\psi_+(u)+4\psi''_+(u)\\
  		&+4({\rm Coth}(u)+{\rm Cot}(u))\psi'_+(u)=0,\\
  	&\big(4(\omega^2L^2{\rm Csc}^2(u)-m^2L^2+mL{\rm Coth}(u))-2{\rm Csch}^2(u)\\
  	&+{\rm Coth}^2(u)+2{\rm Coth}(u){\rm Cot}(u)+\big)\psi_-(u)+4\psi''_-(u)\\
  	&+4({\rm Coth}(u)+{\rm Cot}(u))\psi'_-(u)=0.
  \end{aligned}
    	  \right.
  \end{equation} 
  We solve these two second-order differential equations~\eqref{eq:53}. The two spinor components behave as
  \begin{equation}
  	\begin{aligned}
  \label{eq:54}
  \psi_+&\sim C_1\,_{2}F_1\bigg[-i\omega L,\frac{1}{2}-i(m+\omega)L,\frac{1}{2}-im L;e^{2iu}\bigg]\\
  &\times(1-e^{2iu})^{\frac{1}{2}-i\omega L}+ C_2\,i(1-e^{2iu})^{\frac{1}{2}-i\omega L}e^{iu-(2u+\pi)mL}\\
 &\times\,_{2}F_1\bigg[1-i\omega L,\frac{1}{2}+i(m-\omega)L,\frac{3}{2}+2im L;e^{2iu}\bigg];\\
\end{aligned}
  \end{equation}
  \begin{equation}
  	\begin{aligned}
  \label{eq:55}
  \psi_-&\sim C_1\,_{2}F_1\bigg[-i\omega L,\frac{1}{2}+i(m-\omega)L,\frac{1}{2}+im L;e^{2iu}\bigg]\\
  &\times(1-e^{2iu})^{\frac{1}{2}-i\omega L}+C_2\,i(1-e^{2iu})^{\frac{1}{2}-i\omega L}e^{iu+(2u+\pi) mL}\\
  &\times\,_{2}F_1\bigg[1-i\omega L,\frac{1}{2}-i(\omega+m)L,\frac{3}{2}-im L;e^{2iu}\bigg].\\
    \end{aligned}
  \end{equation}
  Expanding the hypergeometric functions in the solutions~\eqref{eq:46} near the horizon, as $u\rightarrow 0$, one finds the two linearly independent solutions given by: $\psi_+\sim (1-e^{2iu})^{\frac{1}{2}-i\omega L}\big(b_1\psi_{\rm n.}+a_1\psi_{\rm n.n.}\big)$ and $\psi_-\sim (1-e^{2iu})^{\frac{1}{2}-i\omega L}\big(b_2\psi_{\rm n.}+a_2\psi_{\rm n.n.}\big)$. Due to the ingoing boundary condition near the horizon, $\psi_+$ and $\psi_-$ have no term proportional to the positive imaginary frequency index. Then the retarded spinor Green's functions have two sets given as (The details of the hypergeometric function around $u=0$ are also presented in Appendix~\ref{sec:Details2})
  \begin{equation}
  \begin{aligned}
  \label{eq:56}
  &G^R_1(\omega,m)=\frac{b_1}{a_1}\\
  &=-ie^{\pi mL}\frac{\Gamma(\frac{1}{2}-imL)\Gamma(1+i\omega L)\Gamma(\frac{1}{2}+imL+i\omega L)}{\Gamma(\frac{3}{2}+imL)\Gamma(i\omega L)\Gamma(\frac{1}{2}-imL+i\omega L)},\\
  &G^R_2(\omega,m)=\frac{b_2}{a_2}\\
  &=-i e^{-\pi mL}\frac{\Gamma(\frac{1}{2}+imL)\Gamma(1+i\omega L)\Gamma(\frac{1}{2}-imL+i\omega L)}{\Gamma(\frac{3}{2}-imL)\Gamma(i\omega L)\Gamma(\frac{1}{2}+imL+i\omega L)}.
\end{aligned}
\end{equation}
  In~\eqref{eq:53} the equation for $\psi_+$ is related to that of $\psi_-$ by $m\rightarrow -m$, so we could see that the retarded spinor Green's functions $G^R_1(\omega,m)$ and $G^R_2(\omega,m)$ are invariant under $m\rightarrow -m$ from~\eqref{eq:56}. Although the poles and zeros of the two Green's functions are different:
  \begin{equation}
  \label{eq:58}
  \left\{
  \begin{aligned}
  	 &\frac{1}{2}+Lm+i\omega L=0,-1,-2,-3,\cdots\, \\
  	 &\text{\big(poles of $G^R_1(\omega,m)$ and zeros of $G^R_2(\omega,m)$\big)}\,,\\
  	 &\frac{1}{2}-Lm+i\omega L=0,-1,-2,-3,\cdots\,\\
  	 &\text{\big(zeros of $G^R_1(\omega,m)$ and poles of $G^R_2(\omega,m)$\big)}\,,
  \end{aligned}
  \right.
  \end{equation}
  the pole-skipping points are the same:
  \begin{equation}
  \label{eq:59}
  \begin{split}
  	\mathfrak{w}&=\frac{i}{2}, \quad \Delta_2=\frac{1}{2}\,;\\
  	\mathfrak{w}&=\frac{3i}{2}, \quad \Delta_2=\frac{1}{2},\frac{3}{2}\,;\\
  	\mathfrak{w}&=\frac{5i}{2}, \quad \Delta_2=\frac{1}{2},\frac{3}{2},\frac{5}{2}\,;\\
  	\mathfrak{w}&=\frac{7i}{2}, \quad \Delta_2=\frac{1}{2},\frac{3}{2},\frac{5}{2},\frac{7}{2}\,;\\
  	\mathfrak{w}&=\frac{9i}{2}, \quad \Delta_2=\frac{1}{2},\frac{3}{2},\frac{5}{2},\frac{7}{2},\frac{9}{2}\,;\\
  	&\qquad \qquad \vdots
  \end{split}
  \end{equation}
  Where $\Delta_2=\frac{1}{2}+ im$ is the operator dimension of Dirac field mentioned in the Sec~\ref{sec:12}.~\eqref{eq:59} agree with the pole-skipping points obtained from near-horizon behavior~\eqref{eq:30}, as shown in Figure~\ref{fig:Figure2}. 
 We use the method of solving with the bulk retarded Green's function, which produces the same values for the pole-skipping points as those obtained through near-horizon analysis in Sec.~\ref{sec:1}. The leading-order imaginary frequency of $\omega_{dS}=i2\pi T_{dS}(1-s)$ at these special points depends on the $s$-spinor of the bulk field. These locations cause the Green's function to become an indeterminate form of type $0/0$, and this non-uniqueness corresponds to the non-uniqueness of the incoming solution of the bulk equation at the pole-skipping positions near the horizon. There is no clear understanding of the Green's function in dS spacetime; however, we have tried a similar prescription as in AdS spacetime. The whole precise Green's functions may be different from the results in this paper, but the positions of the pole-skipping points of the Green's function in dS spacetime will be the same because the near horizon analysis is still valid regardless of the definition of the Green's function in dS spacetime. 	
	
\section{Pole-skipping in DSSYK model and dS$_2$/DSSYK$_\infty$ correspondence}
\label{sec:3}
 In this section, we will compute the numerical values of the pole-skipping points in DSSYK in the infinite-temperature limit and compare the results with those obtained in the dS$_2$ spacetime. We start with a brief review of the DSSYK model. The $N$ fermions $\psi_{x}\;(x=1,\dots ,N)$ obeying $\{\psi_{x},\psi_{y}\}=2\delta_{xy}$ interact all-to-all in $p$-body groups, and this coupling is random. The Hamiltonian is given as~\cite{Cotler,Berkooz,Berkooz2,Khramtsov}
\begin{equation}
\label{eq:31}
\begin{aligned}
H=i^{p/2}\sum_{1\leq x_1<\dots<x_p\leq N}&J_{x_1\dots x_p}\psi_{x_1}\dots\psi_{x_p}\,,\\
 \langle J^2_{x_1\dots x_p}\rangle=&\frac{\mathcal{J}^2}{\lambda\left(
	\begin{smallmatrix}
		N \\
		p 
	\end{smallmatrix}
	\right)},
	\end{aligned}
\end{equation}
with the $J_{x_1\dots x_p}$ being random couplings. The $\psi_{x}\;(x=1,\dots ,N)$ are Majorana fermions satisfying $\{\psi_{x},\psi_{y}\}=2\delta_{xy}$. This is a large $N$ theory, where $N$ and $p$ are going to be large. In the double scaling limit, the SYK model satisfies
\begin{equation}
\label{eq:32}
N,\,p\rightarrow \infty \quad \quad \lambda\equiv\frac{2p^2}{N}=\text{fixed}.
\end{equation}

\begin{figure}[htp]
	\begin{centering}
		\includegraphics[scale=0.45]{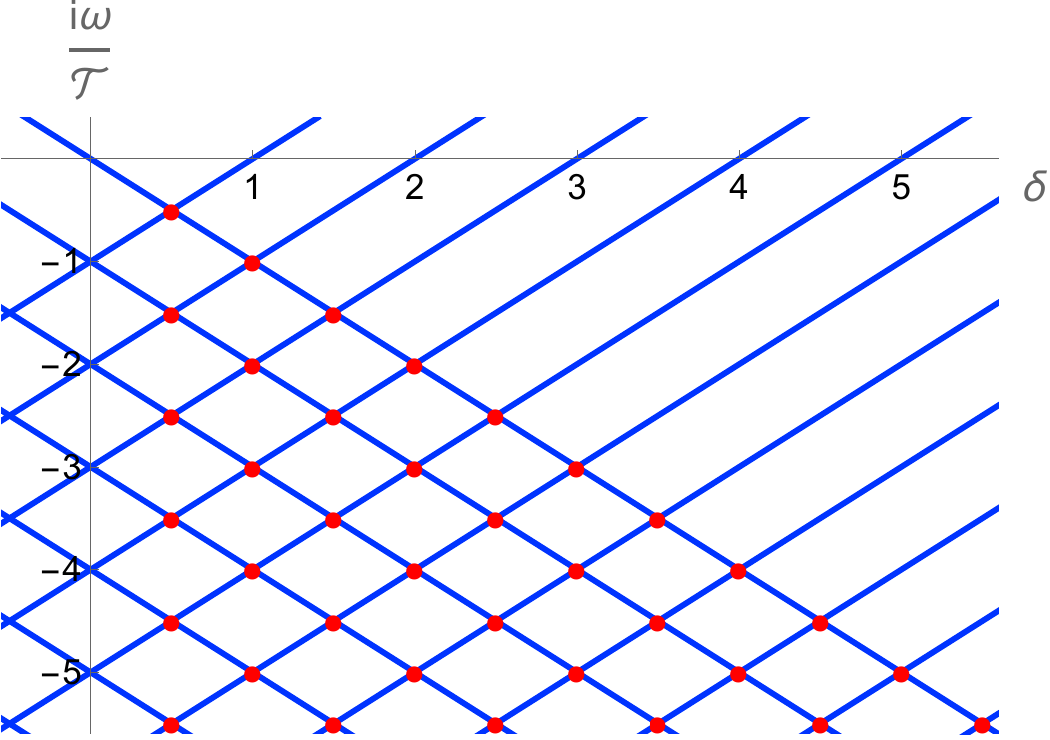}
		\par\end{centering}
	\caption{\label{fig:Figure3} The blue lines are poles and zeros of Green's function in the DSSYK$_\infty$ model, and the red points where they intersect indicate the positions of pole-skipping. ($\delta$ is positive from the definition.)}
\end{figure}

Operators in DSSYK at classical $\lambda\rightarrow 0$ limit are essentially characterized by the number of fermions the operator contains. The normalized two-point function of the same operator is given as~\cite{Narovlansky,Goel}
\begin{equation}
\label{eq:33}
G(t)=\bigg[\frac{{\rm cos}\frac{\pi v}{2}}{{\rm cos}\big(\frac{\pi v}{2}(1-\frac{2it}{\beta})\big)}\bigg]^{2\delta},
\end{equation}
a new parameter $\widetilde{\lambda}$ will be defined. The symbol $\delta=\widetilde{\lambda}/\lambda$ represents the size of the operator, and it can be seen that it is positive. The relationship between parameter $v$ and parameter $\mathcal{J}$ is as follows
\begin{equation}
\label{eq:34}
\beta\mathcal{J}=\frac{\pi v}{{\rm cos}\frac{\pi v}{2}}.
\end{equation}
From relation~\eqref{eq:34}, it can be concluded that in the high-temperature limit as $\beta\rightarrow 0$, the ratio $\frac{\pi v}{\beta}$ approaches $\mathcal{J}$.
In infinite-temperature limit $\beta\mathcal{J}\rightarrow 0$,~\eqref{eq:33} becomes a simple expression
\begin{equation}
\label{eq:new1}
G_{\rm infinite}(t)=\bigg(\frac{1}{{\rm cosh}(\mathcal{J}t)}\bigg)^{2\delta}
\end{equation}
The retarded Green's function in frequency space is defined as~\cite{Weinberg,Parcollet,Qi}
\begin{equation}
\label{eq:45}
G^R(\omega)=-i\int dt\,\theta(t)\,e^{i\omega t}G(t).
\end{equation}
The retarded Green's function becomes
\begin{equation}
\begin{aligned}
\label{eq:36}
&G_{\rm infinite}^R(\omega)=-i\int dt\,\theta(t)\,e^{i\omega t}G_{\rm infinite}(t)\\
&=\frac{-ie^{i\omega t}(1+e^{2\mathcal{J}t})\big({\rm sec}(\mathcal{J}t)\big)^{2\delta}}{2\mathcal{J}\delta+i\omega}\,_2{\bf F}_1\big(a,b;a+b+\mathcal{N};z\big),\\
\end{aligned}
\end{equation}
where $a=1,\;b=1-\delta+\frac{i\omega}{2\mathcal{J}},\;\mathcal{N}=2\delta-1,\;z=-e^{2\mathcal{J}t}$. We now calculate the pole-skipping points of Green's function~\eqref{eq:36}. The function $_2{\bf F}_1\big(a,b;a+b+\mathcal{N};z\big)$ could be expanded into three expressions based on three cases of parameter $\mathcal{N}$~\cite{Ahn3}. However, the pole-skipping positions for these three expressions are the same. We simplify the forms into
\begin{equation}
	\begin{aligned}
\label{eq:37}
&_2{\bf F}_1\big(1,1-\delta+\frac{i\omega}{2\mathcal{J}};1+\delta+\frac{i\omega}{2\mathcal{J}};-e^{2\mathcal{J}t}\big)\\
&=\frac{\Gamma(1+\delta+\frac{i\omega}{2\mathcal{J}})}{\Gamma(\delta+\frac{i\omega}{2\mathcal{J}})}A(\mathcal{N})+\frac{\Gamma(1+\delta+\frac{i\omega}{2\mathcal{J}})}{\Gamma(1-\delta+\frac{i\omega}{2\mathcal{J}})}B(\mathcal{N}).
\end{aligned}
\end{equation}
The values of coefficients $A(\mathcal{N})$ and $B(\mathcal{N})$ are determined by three different cases of $\mathcal{N}$, and only zeros exist in these two coefficients. (The details of these coefficients are presented in Appendix~\ref{sec:Details3}.) The equation~\eqref{eq:36} becomes
\begin{equation}
	\begin{aligned}
\label{eq:38}
G_{\rm infinite}^R(\omega)&=e^{i\omega t}(1+e^{2\mathcal{J}t})\big({\rm sec}(\mathcal{J}t)\big)^{2\delta}\\
&\times\bigg(\frac{\Gamma(1+\delta+\frac{i\omega}{2\mathcal{J}})}{(2\mathcal{J}\delta+i\omega)\Gamma(\delta+\frac{i\omega}{2\mathcal{J}})}A(\mathcal{N})\\
&+\frac{\Gamma(1+\delta+\frac{i\omega}{2\mathcal{J}})}{(2\mathcal{J}\delta+i\omega)\Gamma(1-\delta+\frac{i\omega}{2\mathcal{J}})}B(\mathcal{N})\bigg).
\end{aligned}
\end{equation}
For the term in front of coefficient $A(\mathcal{N})$, the only pole $(2\mathcal{J}\delta+i\omega)$ is reduced by the zero obtained from $\frac{\Gamma(1+\delta+\frac{i\omega}{2\mathcal{J}})}{\Gamma(\delta+\frac{i\omega}{2\mathcal{J}})}=(\delta+\frac{i\omega}{2\mathcal{J}})$. So we focus on the term in front of coefficient $B(\mathcal{N})$, the pole-skipping positions obtained from the intersections of zeros and poles in term $\frac{\Gamma(1+\delta+\frac{i\omega}{2\mathcal{J}})}{(2\mathcal{J}\delta+i\omega)\Gamma(\delta+\frac{i\omega}{2\mathcal{J}})}$. We could obtain these special points
\begin{equation}
\label{eq:65}
\begin{split}
	\omega&=i\mathcal{J}, \quad \delta=\frac{1}{2}\,;\\
	\omega&=2i\mathcal{J}, \quad \delta=1\,;\\
	\omega&=3i\mathcal{J}, \quad \delta=\frac{1}{2},\frac{3}{2}\,;\\
	\omega&=4i\mathcal{J}, \quad \delta=1,2\,;\\
	\omega&=5i\mathcal{J}, \quad \delta=\frac{1}{2},\frac{3}{2},\frac{5}{2}\,;\\
	\omega&=6i\mathcal{J}, \quad \delta=1,2,3\,;\\
	&\qquad \qquad \vdots
\end{split}
\end{equation}
In the DSSYK$_\infty$ model~\cite{Susskind3} defines the ``tomperature'' $\mathcal{T}=2\mathcal{J}$ which remains finite as $\beta\rightarrow 0$. Tomperature $\mathcal{T}$, as an ``effective temperature'', reflects the finiteness of energy and time scale for this system. In the dS metric, the horizon radius $L$ is a single-dimensional parameter. Similarly, the tomperature $\mathcal{T}$ is also a single-dimensional parameter in the DSSYK$_\infty$. To build the duality between two theories, the relation between the two parameters is $\mathcal{T}=1/L$~\cite{Susskind3}. Then,~\eqref{eq:65} could be expressed in a dimensionless form of frequency $\mathfrak{w}=\frac{\omega}{\mathcal{T}}$ and is divided into
\begin{equation}
\label{eq:39}
\begin{split}
	\mathfrak{w}&=i, \quad \delta=1\,;\\
	\mathfrak{w}&=2i, \quad \delta=1,2\,;\\
	\mathfrak{w}&=3i, \quad \delta=1,2,3\,;\\
	\mathfrak{w}&=4i, \quad \delta=1,2,3,4\,;\\
	\mathfrak{w}&=5i, \quad \delta=1,2,3,4,5\,;\\
	&\qquad \qquad \vdots
\end{split}
\end{equation}
and half-integers
\begin{equation}
\label{eq:40}
\begin{split}
	\mathfrak{w}&=\frac{i}{2}, \quad \delta=\frac{1}{2}\,;\\
	\mathfrak{w}&=\frac{3i}{2}, \quad \delta=\frac{1}{2},\frac{3}{2}\,;\\
	\mathfrak{w}&=\frac{5i}{2}, \quad \delta=\frac{1}{2},\frac{3}{2},\frac{5}{2}\,;\\
	\mathfrak{w}&=\frac{7i}{2}, \quad \delta=\frac{1}{2},\frac{3}{2},\frac{5}{2},\frac{7}{2}\,;\\
	\mathfrak{w}&=\frac{9i}{2}, \quad \delta=\frac{1}{2},\frac{3}{2},\frac{5}{2},\frac{7}{2},\frac{9}{2}\,;\\
	&\qquad \qquad \vdots
\end{split}
\end{equation}
~\eqref{eq:39} and~\eqref{eq:40} are similar to the pole-skipping points in dS$_2$ spacetime. We depict those special points in Figure~\ref{fig:Figure3}. From the perspective of the dS$_2$/DSSYK$_\infty$ correspondece, the horizon as the boundary of the static patch of de Sitter spacetime, and the pole-skipping structure at this area is consistent with that of DSSYK in the infinite-temperature limit. 
We could see that~\eqref{eq:39} corresponds one-to-one with~\eqref{eq:18} and~\eqref{eq:50}, while~\eqref{eq:40} corresponds one-to-one with~\eqref{eq:30} and~\eqref{eq:59}. There are only the scalar field and fermionic field in dS$_2$ spacetime, and they correspond to the integer part and the half-integer part of the pole-skipping points, respectively. We could see that the pole-skipping points in DSSYK$_\infty$ in the classical limit include the sum of the pole-skipping points of these two types of bulk fields in the dS$_2$ spacetime.\\
\indent We will discuss the relationship between the DSSYK model and the geometry of dS spacetime. Ref.~\cite{Goel} provides a insight into the DSSYK and the curvature of two-dimensional spacetime. It presents a Liouville form DSSYK model
\begin{equation}
\label{eq:66}
I=\frac{N}{16 p^2}\int d\tau d r\big[-\frac{1}{2}(\partial_{\tau}g)^2+\frac{1}{2}(\partial_r g)^2-2\mathcal{J}^2 e^{g(\tau,r)}\big].
\end{equation}
Liouville theory describes the quantum mechanics of the Weyl factor with respect to the naive metric appearing in the kinetic term, so when the Liouville equations of motion are imposed on $g$, it could take the form of a two-dimensional de Sitter spacetime~\cite{Goel}
\begin{equation}
\label{eq:67}
ds^2=\frac{{\rm cos}(\frac{\pi v}{2})^2}{{\rm cos}\big(\frac{\pi v}{2}(1-\frac{2it}{\beta})\big)^2}\big(dt^2+dr^2\big).
\end{equation}
In the case of infinite-temperature $v\rightarrow 0$, the metric~\eqref{eq:67} becomes
\begin{equation}
\label{eq:68}
ds^2=\frac{1}{{\rm cosh}(\mathcal{J}t)^2}\big(dt^2+dr^2\big).
\end{equation}
We could obtain Ricci scalar form from~\eqref{eq:68}
\begin{equation}
\label{eq:69}
R=2\mathcal{J}^2,
\end{equation}
which is a constant positive curvature. The DSSYK$_\infty$ model could provide two-dimensional dS geometry; therefore, it preserves the same pole-skipping structure in the static patch of dS$_2$ spacetime. We consider that this consistency in the numerical values of these uncertain points could provide a conjecture for the dS$_2$/DSSYK$_\infty$ correspondence. 

Ref.~\cite{Goel} also demonstrates that the Liouville form of the double-scaled SYK model naturally gives rise to a two-dimensional AdS spacetime structure. A feature of the DS/dS correspondence is that both dS$_{d+1}$ and AdS$_{d+1}$ spacetimes can be foliated into dS$_{d}$ slices. This correspondence establishes a duality between ${d+1}-$dimensional quantum gravity and the field theory defined on these dS$_{d}$ slices. If we regard the DSSYK model as a field theory on the dS$_{1}$ slice, it can render both AdS$_2$ spacetime forms and dS$_2$ spacetime forms, which could be considered as a signature of the DS/dS correspondence. 

\section{Conclusion}
\label{sec:Conclusion}

In summary, we develop the pole-skipping structure at the cosmic horizon $r=L$ in the static patch of dS spacetime. For fields with spin $s$, their leading pole-skipping frequencies satisfy $\omega_{dS}=i2\pi T_{dS}(1-s)$, which is exactly opposite to the frequencies in AdS spacetime. 

We also obtain a numerical consistent relationship between the static patch of dS$_2$ spacetime and the DSSYK$_\infty$ model. The pole-skipping points are regarded as a series of uncertain positions within the system. These uncertain points directly correspond to the integer pole-skipping for the scalar field and the half-integer pole-skipping for the fermionic field in dS$_2$ gravity, respectively. From equations~\eqref{eq:17} and~\eqref{eq:30}, we can see that in dS$_2$ spacetime, each order of pole-skipping points for the scalar field and fermionic field has a position with mass equal to $0$, while the mass squared of the subsequent pole-skipping points is negative. These massless pole-skipping points have physical meaning. Although the other pole-skipping points with negative mass squared are non-physical, they correspond one-to-one with the pole-skipping points in the DSSYK$_\infty$ model. We list these special points that have the mathematical form of the pole-skipping structure to provide a reference for the dS$_2$/DSSYK$_\infty$ correspondence.
  
The correspondence between dS$_2$ and DSSYK$_\infty$ can be viewed as a correspondence between the quantum gravity of static patch in dS$_2$ and the field theory which is unitary on the dS$_1$ slice. We demonstrate the pole-skipping phenomenon in the application of the DS/dS correspondence. We believe our results highlight the universal role of pole-skipping as a diagnostic tool for holographic duality and could provide a new perspective for the AdS/CFT and the DS/dS correspondence. 

\section*{Acknowledgements}

This work is supported by the National Natural Science Foundation of China (No.12275166), Natural Science Foundation of Henan Province of China (No.252300420892), and the Launching Funding of Henan University of Technology (No.31401598). This work was also supported by the Basic Science Research Program through the National Research Foundation of Korea (NRF) funded by the Ministry of Science, ICT and Future Planning (NRF- 2021R1A2C1006791), the GIST Research Institute (GRI) and the AI-based GIST Research Scientist Project grant funded by the GIST in 2023.

\appendix

\section{Details of near-horizon expansions} 
\label{sec:Details1}
In this appendix, we show the details of the near-horizon expansions of the equations of motion. We can calculate a Taylor series solution to the scalar mode $\Psi(r)$ equation of motion when the matrix equation~\eqref{eq:15} is satisfied. The first few elements of this matrix are shown below
\begin{equation}
		\begin{aligned}
M_{11}=\frac{m^2}{2};&\\
M_{21}=0,\quad &M_{22}=\frac{m^2-f''(r_h)}{4};\\
M_{31}=0,\quad &M_{32}=-\frac{f^{(3)}(r_h)}{12},\quad M_{33}=\frac{m^2-3f''(r_h)}{6};\\
M_{41}=0,\quad &M_{42}=-\frac{f^{(4)}(r_h)}{48},\quad M_{43}=-\frac{f^{(3)}(r_h)}{6},\\
&M_{44}=\frac{m^2-6f''(r_h)}{8};\\
M_{51}=0,\quad &M_{52}=-\frac{f^{(5)}(r_h)}{240},\quad M_{53}=-\frac{f^{(4)}(r_h)}{24},\\
&M_{54}=-\frac{f^{(3)}(r_h)}{4},\quad M_{55}=\frac{m^2-10f''(r_h)}{10};\\
\qquad \vdots
	\end{aligned}
\end{equation}

\section{Expressions for the hypergeometric function $_2{\bf F}_1\big(a,\,b;\,a+b+\mathcal{N};\,1\big)$}
\label{sec:Details2}
In this appendix, we consider four cases depending on whether $\mathcal{N}$ is a non-integer, a positive integer, zero, or a negative integer. As $z \rightarrow 1^{-}$, the hypergeometric function $_2{\bf F}_1\big(a,\,b;\,a+b+\mathcal{N};\,z\big)$ becomes~\cite{Ahn3}:
\\\\
{$\bullet$ $\mathcal{N}$ is a non-integer ($\mathcal{N}\rightarrow n$)}:
\begin{equation}
_2{\bf F}_1\big(a,b;a+b+n;z\big)\overset{z\rightarrow 1^{-}}{\simeq}\frac{\Gamma(a+b+n)\Gamma(n)}{\Gamma(a+n)\Gamma(b+n)}.
\end{equation}
\\
{$\bullet$ $\mathcal{N}$ is a positive integer ($\mathcal{N}\rightarrow N$)}:
\begin{equation}
_2{\bf F}_1\big(a,b;a+b+N;z\big)\overset{z\rightarrow 1^{-}}{\simeq}\frac{\Gamma(a+b+N)\Gamma(N)}{\Gamma(a+N)\Gamma(b+N)}.
\end{equation}
\\
{$\bullet$ $\mathcal{N}$ is zero}:
\begin{equation}
_2{\bf F}_1\big(a,b;a+b;z\big)\overset{z\rightarrow 1^{-}}{\simeq}-\frac{\Gamma(a+b)}{\Gamma(a)\Gamma(b)}[\psi(a)+\psi(b)].
\end{equation}
\\
{$\bullet$ $\mathcal{N}$ is a negative integer ($\mathcal{N}\rightarrow -N$)}:
\begin{equation}
	\begin{aligned}
_2{\bf F}_1\big(a,b;a+b-N;z\big)\overset{z\rightarrow 1^{-}}{\simeq}&-\frac{(-1)^N}{N!}\frac{\Gamma(a+b-N)}{\Gamma(a-N)\Gamma(b-N)}\\
&\times[\psi(a)+\psi(b)],
	\end{aligned}
\end{equation}
where $\psi(a)=\Gamma'(a)/\Gamma(a)=d\,{\rm ln}\Gamma(a)/da$.

\section{Expressions for the hypergeometric function $_2{\bf F}_1\big(a,\,b;\,a+b+\mathcal{N};\,z\big)$} 
\label{sec:Details3}
In this appendix, we list the details of equation~\eqref{eq:36}. The hypergeometric function $_2{\bf F}_1\big(1,1-\delta+\frac{i\omega}{2\mathcal{J}};1+\delta+\frac{i\omega}{2\mathcal{J}};-e^{2\mathcal{J}t}\big)$ can be written in the form of $_2{\bf F}_1\big(a,\,b;\,a+b+\mathcal{N};\,z\big)$, where $a=1,\;b=1-\delta+\frac{i\omega}{2\mathcal{J}},\;\mathcal{N}=2\delta-1,\;z=-e^{2\mathcal{J}t}$. There are four outcomes when $\mathcal{N}$ is a non-integer, positive integer, zero, or a negative integer~\cite{Ahn3}:
\\\\
{$\bullet$ $\mathcal{N}$ is a non-integer ($\mathcal{N}\rightarrow n$)}:
\begin{widetext}
	\begin{equation}
		\begin{aligned}
_2{\bf F}_1\big(a,b;a+b+n;z\big)=&\frac{\Gamma(a+b+n)\Gamma(n)}{\Gamma(a+n)\Gamma(b+n)}\sum^{\infty}_{k=0}\frac{(a)_k (b)_k}{(1-n)_k k!}(1-z)^k+(1-z)^n\frac{\Gamma(a+b+n)\Gamma(-n)}{\Gamma(a)\Gamma(b)}\\
&\times \sum^{\infty}_{k=0}\frac{(a+n)_k(b+n)_k}{(1+n)_k k!}(1-z)^k,
	\end{aligned}
	\end{equation}
\end{widetext}
where $(a)_k=\Gamma(a+k)/\Gamma(a)=a(a+1)\cdots(a+k-1)$.

{$\bullet$ $\mathcal{N}$ is a positive integer ($\mathcal{N}\rightarrow N$)}:
\begin{widetext}
\begin{equation}
	\begin{aligned}
&_2{\bf F}_1\big(a,b;a+b+N;z\big)=\frac{\Gamma(a+b+N)}{\Gamma(a+N)\Gamma(b+N)}\sum^{N-1}_{k=0}\frac{(a)_k (b)_k (N-k-1)!}{k!}(z-1)^k-(z-1)^N\frac{\Gamma(a+b+N)}{\Gamma(a)\Gamma(b)}\\
&\times\sum^{\infty}_{k=0}\frac{(a+N)_k(b+N)_k}{k!(k+N)!}(1-z)^k[{\rm log}(1-z)-\psi(k+1)-\psi(k+N+1)+\psi(a+k+N)+\psi(b+k+N)]
	\end{aligned}
\end{equation}
\end{widetext}

{$\bullet$ $\mathcal{N}$ is zero}:
\begin{widetext}
\begin{equation}
		\begin{aligned}
_2{\bf F}_1\big(a,b;a+b;z\big)=-\frac{\Gamma(a+b)}{\Gamma(a)\Gamma(b)}\sum^{\infty}_{k=0}\frac{(a)_k (b)_k}{(k!)^2}(1-z)^k[{\rm log}(1-z)-2\psi(k+1)+\psi(a+k)+\psi(b+k)]
	\end{aligned}
\end{equation}
\end{widetext}
{$\bullet$ $\mathcal{N}$ is a negative integer ($\mathcal{N}\rightarrow -N$)}:
\begin{widetext}
\begin{equation}
		\begin{aligned}
&_2{\bf F}_1\big(a,b;a+b-N;z\big)=\frac{\Gamma(a+b-N)}{\Gamma(a)\Gamma(b)}\sum^{N-1}_{k=0}\frac{(a-N)_k (b-N)_k (N-k-1)!}{k!}(1-z)^{-N}(z-1)^k\\
&-(-1)^N\frac{\Gamma(a+b-N)}{\Gamma(a-N)\Gamma(b-N)}\sum^{\infty}_{k=0}\frac{(a)_k(b)_k}{k!(k+N)!}(1-z)^k\times[{\rm log}(1-z)-\psi(k+1)-\psi(k+N+1)+\psi(a+k)+\psi(b+k)].
	\end{aligned}
\end{equation}
\end{widetext}

\section{SYK model in the low-temperature limit}
 In this appendix, we consider the pole-skipping points in the low-temperature limit of the SYK model, which exhibits a correspondence with the results from JT gravity in two-dimensional AdS. The conclusions obtained are the same as those we previously obtained in Ref.~\cite{Yuan3}. 
 
\begin{figure}[htp]
	\begin{centering}
		\includegraphics[scale=0.45]{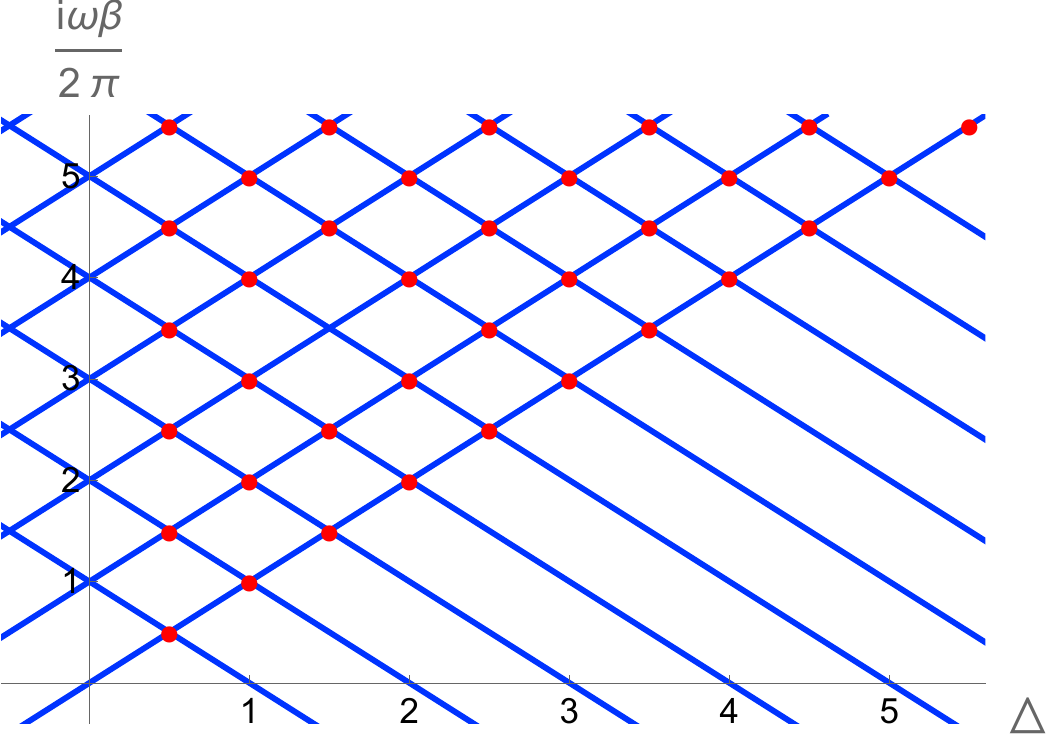}
		\par\end{centering}
	\caption{\label{fig:Figure4} The blue lines are poles and zeros of Green's function in the low-temperature SYK model, and the red points where they intersect indicate the positions of pole-skipping. ($\Delta$ is positive from the definition.)}
\end{figure}

 The SYK model develops an emergent conformal symmetry as $\beta\mathcal{J}\rightarrow \infty$, and the two-point function is given as~\cite{Stanford,Polchinski,Roberts,Tarnopolsky}
\begin{equation}
	\label{eq:new2}
	G_c(t)=b\bigg(\frac{\pi}{\beta {\rm sinh}\frac{\pi t}{\beta}}\bigg)^{2\Delta}. 
\end{equation}
The retarded Green's function is
\begin{equation}
	\begin{aligned}
		\label{eq:new3}
		G^R_{\rm low}(\omega)=&-i\int dt\,\theta(t)\,e^{i\omega t}G_{c}(t)\\
		=&\frac{ib\beta\big(-\frac{i\pi{\rm csch}(\frac{\pi t}{\beta})}{\beta}\big)^{2\Delta}e^{i\omega t}(1-e^{-\frac{2\pi t}{\beta}})}{2\pi\Delta-i\beta\omega}\\
		&\times_2{\bf F}_1\big(a,b;a+b+\mathcal{N};z\big),\\
	\end{aligned}
\end{equation}
where $a=1,\;b=1-\delta-\frac{i\beta\omega}{2\pi},\;\mathcal{N}=2\delta-1,\;z=-e^{\frac{2\pi t}{\beta}}$. Using our method in the infinite-temperature limit, we can obtain the pole-skipping points at low-temperature
\begin{equation}
	\label{eq:new4}
	\begin{split}
		\omega&=-\frac{i\pi}{\beta}, \quad \Delta=\frac{1}{2}\,;\\
		\omega&=-\frac{i2\pi}{\beta}, \quad \Delta=1\,;\\
		\omega&=-\frac{i3\pi}{\beta}, \quad \Delta=\frac{1}{2},\frac{3}{2}\,;\\
		\omega&=-\frac{i4\pi}{\beta}, \quad \Delta=1,2\,;\\
		\omega&=-\frac{i5\pi}{\beta}, \quad \Delta=\frac{1}{2},\frac{3}{2},\frac{5}{2}\,;\\
		\omega&=-\frac{i6\pi}{\beta}, \quad \Delta=1,2,3\,;\\
		&\qquad \qquad \vdots
	\end{split}
\end{equation}
After rescaling the frequency $\mathfrak{w}=\frac{\beta\omega}{2\pi}$ , it is evident that the integer points
\begin{equation}
	\label{eq:new5}
	\begin{split}
		\mathfrak{w}&=-i, \quad \Delta=1\,;\\
		\mathfrak{w}&=-2i, \quad \Delta=1,2\,;\\
		\mathfrak{w}&=-3i, \quad \Delta=1,2,3\,;\\
		\mathfrak{w}&=-4i, \quad \Delta=1,2,3,4\,;\\
		\mathfrak{w}&=-5i, \quad \Delta=1,2,3,4,5\,;\\
		&\qquad \qquad \vdots
	\end{split}
\end{equation}
and the half-integer points 
\begin{equation}
	\label{eq:new6}
	\begin{split}
		\mathfrak{w}&=-\frac{i}{2}, \quad \Delta=\frac{1}{2}\,;\\
		\mathfrak{w}&=-\frac{3i}{2}, \quad \Delta=\frac{1}{2},\frac{3}{2}\,;\\
		\mathfrak{w}&=-\frac{5i}{2}, \quad \Delta=\frac{1}{2},\frac{3}{2},\frac{5}{2}\,;\\
		\mathfrak{w}&=-\frac{7i}{2}, \quad \Delta=\frac{1}{2},\frac{3}{2},\frac{5}{2},\frac{7}{2}\,;\\
		\mathfrak{w}&=-\frac{9i}{2}, \quad \Delta=\frac{1}{2},\frac{3}{2},\frac{5}{2},\frac{7}{2},\frac{9}{2}\,;\\
		&\qquad \qquad \vdots
	\end{split}
\end{equation}
We depict those in Figure~\ref{fig:Figure4}.
In the low-temperature limit, the correlation functions of SYK are consistent with the correlators obtained in the Schwarzian theory, which describes the dynamics of JT gravity in two-dimensional AdS spacetime. The numerical results~\eqref{eq:new5} and~\eqref{eq:new6} of the pole-skipping that we obtained in the low-temperature SYK model correspond well to the results of the scalar field and the fermionic field in two-dimensional AdS spacetime, respectively. We hope it can help study the duality mechanism between quantum gravity in two-dimensional AdS and dS spacetimes and one-dimensional gauge theory.

\begin{thebibliography}{99}
	\bibitem{Hooft} G. ’t Hooft, {Dimensional reduction in quantum gravity}, Conf. Proc. C {\bf 930308}, 284 (1993), arXiv:gr-qc/9310026.
	\bibitem{Susskind5} L. Susskind, {The World as a hologram}, J. Math. Phys. {\bf 36} 6377 (1995), arXiv:hep-th/9409089.
	\bibitem{Maldacena2} J. Maldacena, {The Large-N Limit of Superconformal Field Theories and Supergravity}, Int. J. Theor. Phys. {\bf 38}, 1113 (1999), arXiv:9711200.
	\bibitem{gubser} S. S. Gubser, I. R. Klebanov and A. M. Polyakov, {Gauge theory correlators from noncritical string theory}, Phys. Lett. B. {\bf428}, 105 (1998), arXiv:9802109.
	\bibitem{witten1} E. Witten, {Anti-de Sitter space and holography}, Adv. Theor. Math. Phys. {\bf 2}, 253 (1988), arXiv:9802150.
	\bibitem{witten2} E. Witten, {Anti-de Sitter space, thermal phase transition, and confinement in gauge theories}, Adv. Theor. Math. Phys. {\bf 2}, 505 (1998), arXiv:9803131.
	\bibitem{Gibbons} G. W. Gibbons and S. W. Hawking, {Cosmological Event Horizons, Thermodynamics, And Particle Creation}, Phys. Rev. D {\bf 15}, 2738 (1977).
	\bibitem{Maldacena4} J. M. Maldacena, {Non-Gaussian features of primordial fluctuations in single field inflationary models}, J. High Energ. Phys. {\bf 05}, 013 (2003), arXiv:astro-ph/0210603.
	\bibitem{Strominger} A. Strominger, {The dS/CFT correspondence}, J. High Energ. Phys. {\bf 10}, 034 (2001), arXiv:hep-th/0106113.
	\bibitem{Witten3} E. Witten, {Quantum gravity in de Sitter space}, in Strings 2001: International Conference, 6, 2001, arXiv:hep-th/0106109.
	\bibitem{Strominger2} M. Spradlin, A. Strominger, and A. Volovich, {Les Houches lectures on de Sitter space}, arXiv:hep-th/0110007.
	\bibitem{Klemm} D. Klemm, {Some aspects of the de Sitter/CFT correspondence}, Nucl. Phys. B {\bf 625}, 295–311 (2002), arXiv:hep-th/0106247.
	\bibitem{Harlow} D.~Harlow and D.~Stanford, {Operator Dictionaries and Wave Functions in AdS/CFT and dS/CFT}, arXiv:1104.2621.
	\bibitem{Strominger3} D. Anninos, T. Hartman and A. Strominger, {Higher spin realization of the dS/CFT correspondence}, Class. Quant. Grav. {\bf 34}, 015009 (2017), arXiv:1108.5735.
	\bibitem{Strominger4} G. S. Ng and A. Strominger, {State/Operator Correspondence in Higher-Spin dS/CFT}, Class. Quant. Grav. {\bf 30}, 104002  (2013), arXiv:1204.1057.
	\bibitem{Hikida} Y. Hikida, T. Nishioka, T. Takayanagi, and Y. Taki, {Holography in de Sitter Space via Chern-Simons Gauge Theory}, Phys. Rev. Lett. {\bf 129}, 041601 (2022), arXiv:2110.03197.
	\bibitem{Hikida2} Y. Hikida, T. Nishioka, T. Takayanagi, and Y. Taki, {CFT duals of three-dimensional de Sitter gravity}, J. High Energ. Phys. {\bf 2022}, 129 (2022), 	arXiv:2203.02852.
	\bibitem{Chen} H. Y. Chen and Y. Hikida, {Three-Dimensional de Sitter Holography and Bulk Correlators at Late Time}, Phys. Rev. Lett. {\bf 129}, 061601 (2022), arXiv:2204.04871.
	\bibitem{Karch} A. Karch, {Auto-localization in de-Sitter space}, J. High Energ. Phys. {\bf 2003}, 050 (2003), arXiv:hep-th/0305192.
	\bibitem{Alishahiha} M. Alishahiha, A. Karch, E. Silverstein, and D. Tong, {The dS/dS Correspondence}, AIP Conf. Proc. {\bf 743}, 393–409 (2004), arXiv:hep-th/0407125.
	\bibitem{Geng1} H. Geng, S. Grieninger, and A. Karch, {Entropy, entanglement and swampland bounds in DS/dS}, J. High Energ. Phys. {\bf 2019}, 105 (2019), arXiv:1904.02170.
	\bibitem{Geng2} H. Geng, {Some information theoretic aspects of de-Sitter holography}, J. High Energ. Phys. {\bf 2020}, 5 (2020), arXiv:1911.02644.
	\bibitem{Geng3} H. Geng, {Non-local entanglement and fast scrambling in de-Sitter holography}, Annals of Physics {\bf 426}, 168402  (2021), arXiv:2005.00021.
	\bibitem{Geng4} H. Geng, Y. Nomura, and H. Y. Sun, {Information paradox and its resolution in de Sitter holography}, Phys. Rev. D {\bf 103}, 126004 (2021), arXiv:2103.07477.
	\bibitem{Grozdanov1} S. Grozdanov, K. Schalm, and V. Scopelliti, {Black Hole Scrambling from Hydrodynamics}, Phys. Rev. Lett. {\bf 120}, 231601 (2018), arXiv:1710.00921.
	\bibitem{Blake1} M. Blake, H. Lee, and H. Liu, {A quantum hydrodynamical description for scrambling and many-body chaos}, J. High Energ. Phys. {\bf 2018} 127 (2018), arXiv:1801.00010.
	\bibitem{Blake2} M. Blake, R. A. Davions, S. Grozdanov, and H. Liu, {Many-body chaos and energy dynamics in holography}, J. High Energ. Phys. {\bf2018}, 35 (2018), arXiv:1809.01169.
	\bibitem{Grozdanov2} S. Grozdanov, {On the connection between hydrodynamics and quantum chaos in holographic theories with stringy corrections}, J. High Energ. Phys. {\bf2019}, 48 (2019), arXiv:1811.09641.
	\bibitem{Das} S. Das, B. Ezhuthachan and A. Kundu, {Real time dynamics from low point correlators in 2d BCFT}, J. High Energ. Phys. {\bf 2019} 141 (2019), arXiv:1907.08763.
	\bibitem{Makoto1} M. Natsuume and T. Okamura, {Holographic chaos, pole-skipping, and regularity}, Progress of Theoretical and Experimental Physics {\bf 1} (2020) 013B07, arXiv:1905.12014.
	\bibitem{Makoto2} M. Natsuume and T. Okamura, {Nonuniqueness of Green's functions at special points}, arXiv:1905.12015.
	\bibitem{BlakeDavison} M. Blake, R. A. Davison, and D. Vegh, {Horizon constraints on holographic Green's functions}, J. High Energ. Phys. {\bf2020}, 77 (2020), arXiv:1904.12883.
	\bibitem{Abbasi1} N. Abbasi and S. Tahery, {Complexified quasinormal modes and the pole-skipping in a holographic system at finite chemical potential}, J. High Energ. Phys. {\bf 2020} 76 (2020), arXiv:2007.10024.
	\bibitem{Abbasi2} N. Abbasi and J. Tabatabaei, {Quantum chaos, pole-skipping and hydrodynamics in a holographic system with chiral anomaly}, J. High Energ. Phys. {\bf2020}, 50 (2020), arXiv:1910.13696.
	\bibitem{Abbasi3} N. Abbasi, and M. Kaminski, {Constraints on quasinormal modes and bounds for critical points from pole-skipping}, J. High Energ. Phys. {\bf 2021}, 265 (2021), arXiv:2012.15820.
	\bibitem{Choi} C. Choi, M. Mezei and G. S{\'a}rosi, {Pole skipping away from maximal chaos}, J. High Energ. Phys. {\bf2021}, 207 (2021), arXiv: 2010.08558.
	\bibitem{Karunava} K. Sil, {Pole skipping and chaos in anisotropic plasma: a holographic study}, J. High Energ. Phys. {\bf2021}, 232 (2021), arXiv:2012.07710.
	\bibitem{Ahn1} Y.~Ahn, V.~Jahnke, H.~S.~Jeong, K.~Y.~Kim,  K.~S.~Lee, and M.~Nishida, {Classifying pole-skipping points}, arXiv:2010.16166.
	\bibitem{Mahdi} M. Atashi and K. Bitaghsir Fadafan, {Holographic pole-skipping of flavor branes}, Journal of Holography Applications in Physics {\bf2}(2), pp. 39-46 (2022). doi: 10.22128/jhap.2022.519.1020.
	\bibitem{Makoto3} M. Natsuume and T. Okamura, {Pole-skipping with finite-coupling corrections}, Phys. Rev. D {\bf100}, 126012 (2019), arXiv:1909.09168.
	\bibitem{N1} N. {\'C}eplak, K. Ramdial, and D. Vegh, {Fermionic pole-skipping in holography}, J. High Energ. Phys. {\bf2020}, 203 (2020), arXiv:1910.02975.
	\bibitem{Yuan1} H. Yuan and X. H. Ge, {Pole-skipping and hydrodynamic analysis in Lifshitz, AdS2 and Rindler geometries}, J. High Energ. Phys. {\bf2021}, 165 (2021), arXiv:2012.15396.
	\bibitem{N2} N. {\'C}eplak and D. Vegh, {Pole skipping and Rarita-Schwinger fields}, Phys. Rev. D {\bf103}, 106009 (2021), arXiv:2101.01490.
	\bibitem{Yuan2} H. Yuan and X. H. Ge, {Analogue of the pole-skipping phenomenon in acoustic black holes}, Eur. Phys. J. C {\bf82}, 167 (2022), arXiv:2110.08074.
	\bibitem{Diandian} D. Wang and Z. Y. Wang, {Pole Skipping in Holographic Theories with Bosonic Fields}, Phys. Rev. Lett. {\bf129}, 231603 (2022), arXiv:2208.01047.
	\bibitem{Jeong} H. S. Jeong, K. Y. Kim, and Y. W. Sun, {Bound of diffusion constants from pole-skipping points: spontaneous symmetry breaking and magnetic field}, J. High Energ. Phys. {\bf 2021}, 105 (2021), arXiv:2104.13084.
	\bibitem{Ning} S. Ning, D. Wang, and Z. Y. Wang, {Pole skipping in holographic theories with gauge and fermionic fields}, J. High Energ. Phys. {\bf 2023}, 84 (2023), arXiv:2308.08191.
	\bibitem{Ahn2} Y. Ahn, V. Jahnke, H. S. Jeong, C. W. Ji, K. Y. Kim, and M. Nishida, {On pole-skipping with gauge-invariant variables in holographic axion theories}, J. High Energ. Phys. {\bf 2024}, 20 (2024), arXiv:2402.12951.
	\bibitem{Grozdanov3} S. Grozdanov and M. Vrbica, {Pole-skipping of gravitational waves in the backgrounds of four-dimensional massive black holes}, Eur. Phys. J. C {\bf 83}, 1103 (2023), arXiv:2303.15921.
	\bibitem{Sachdev} S. Sachdev and J. Ye, {Gapless spin-fluid ground state in a random quantum Heisenberg magnet}, Physical Review Letters 70 3339 (1993), arXiv:cond-mat/9212030.
	\bibitem{Kitaev} A. Kitaev, {A simple model of quantum holography},\\
	http://online.kitp.ucsb.edu/online/entangled15/kitaev/,\\
	http://online.kitp.ucsb.edu/online/entangled15/kitaev2/. Talks at KITP, April 7, 2015 and May 27, 2015.
	\bibitem{Jackiw} R. Jackiw, {Lower dimensional gravity}, Nucl. Phys. B {\bf252}, 343 (1985).
	\bibitem{Teitelboim} C. Teitelboim, {Gravitation and Hamiltonian structure in two space-time dimensions}, Phys. Lett. B {\bf126}, 41 (1983).
	\bibitem{Sarosi} G. {\'S}arosi, AdS2 holography and the SYK model, arXiv:1711.08482.
	\bibitem{Stanford} J. Maldacena and D. Stanford, {Remarks on the Sachdev-Ye-Kitaev model}, Phys. Rev. D {\bf94}, 106002 (2016), arXiv:1604.07818.
	\bibitem{Polchinski} J. Polchinski and V. Rosenhaus, {The Spectrum in the Sachdev-Ye-Kitaev Model}, J. High Energ. Phys. {\bf1604}, 001 (2016), arXiv:1601.06768.
	\bibitem{Roberts} D.A. Roberts, D. Stanford, and A. Streicher, {Operator growth in the SYK model}, J. High Energ. Phys. 2018, 122 (2018),	arXiv:1802.02633.
	\bibitem{Tarnopolsky} G. Tarnopolsky,  {Large q expansion in the Sachdev-Ye-Kitaev model}, Phys. Rev. D 99, 026010 (2019), arXiv:1801.06871.
	\bibitem{Cai1} W. Cai, X. H. Ge, and G. H. Yang, {Diffusion in higher dimensional SYK model with complex fermions}, J. High Energ. Phys. {\bf 2018}, 76 (2018), arXiv:1711.07903.
	\bibitem{Cai2} W. Cai, S. Cao, X. H. Ge, M. Matsumoto, and S. J. Sin, {Non-Hermitian quantum system generated from two coupled Sachdev-Ye-Kitaev models}, Phys. Rev. D {\bf 106}, 106010 (2022), arXiv:2208.10800.
	\bibitem{Cao} J. C. Louw, S. Cao, and X. H. Ge, {Matching partition functions of deformed Jackiw-Teitelboim gravity and the complex SYK model}, Phys. Rev. D {\bf 108}, 086014 (2023), arXiv:2305.05394.
	\bibitem{Cai3} C. Zhang and W. Cai, {$T\bar{T}$ deformation on non-Hermitian two coupled SYK model}, arXiv:2312.03433.
	\bibitem{Cao2} S. Cao and X. H. Ge, {Excitation transmission through a non-Hermitian traversable wormhole}, Phys. Rev. D {\bf 110}, 046022 (2024), 	arXiv:2404.11436.
	\bibitem{Yuan3} H. Yuan, X. H. Ge, K. Y. Kim, C. W. Ji, and Y. Ahn {Pole-skipping points in 2D gravity and SYK model}, J. High Energ. Phys. {\bf 2023}, 157 (2023), arXiv:2303.04801. 
	\bibitem{Cotler} J. S. Cotler, G. Gur-Ari, M. Hanada, J. Polchinski, P. Saad, S. H. Shenker, D. Stanford, A. Streicher, and M. Tezuka, {Black Holes and Random Matrices}, J. High Energ. Phys. \textbf{2017}, 118 (2017), arXiv:1611.04650.
	\bibitem{Berkooz} M. Berkooz, M. Isachenkov, V. Narovlansky, and G. Torrents, {Towards a full solution of the large N double-scaled SYK model}, J. High Energ. Phys. \textbf{2019}, 079 (2019), arXiv:1811.02584.
	\bibitem{Berkooz2} M. Berkooz, V. Narovlansky, and H. Raj, {Complex Sachdev-Ye-Kitaev model in the double scaling limit}, J. High Energ. Phys. \textbf{2021}, 113 (2021), arXiv:2006.13983.
	\bibitem{Khramtsov} M. Khramtsov and E. Lanina, {Spectral form factor in the double-scaled SYK model}, J. High Energ. Phys. \textbf{2021}, 31 (2021), arXiv:2011.01906.
	\bibitem{Lin} H. W. Lin, D. Stanford, {A symmetry algebra in double-scaled SYK}, SciPost Phys. \textbf{15}, 234 (2023), arXiv:2307.15725.
	\bibitem{Berkooz3} M. Berkooz, R. Frumkin, O. Mamroud, and J. Seitz,  {Twisted times, the Schwarzian and its deformations in DSSYK}, arXiv:2412.14238.
	\bibitem{Saad} P. Saad, S. H. Shenker, and D. Stanford, {JT gravity as a matrix integral}, arXiv:1903.11115.
	\bibitem{Maldacena3} J. Maldacena, G. J. Turiaci and Z. Yang, {Two dimensional Nearly de Sitter gravity}, J. High Energ. Phys. {\bf 01}, 139 (2021), arXiv:1904.01911.
	\bibitem{Cotler2} J. Cotler, K. Jensen and A. Maloney, {Low-dimensional de Sitter quantum gravity}, J. High Energ. Phys. {\bf 06}, 048 (2020), arXiv:1905.03780.
	\bibitem{Susskind6} L. Susskind, {Black Holes Hint Towards De Sitter-Matrix Theory}, arXiv:2109.01322.
	\bibitem{Cotler3} J. Cotler and K. Jensen, {Isometric evolution in de Sitter quantum gravity}, arXiv:2302.06603.
	\bibitem{Susskind1} L. Susskind, {Entanglement and Chaos in De Sitter Space Holography: An SYK Example}, JHAP \textbf{1}, no.1, 1-22 (2021), arXiv:2109.14104.
	\bibitem{Susskind2} L. Susskind, {Scrambling in Double-Scaled SYK and De Sitter Space}, arXiv:2205.00315.
	\bibitem{Susskind3} H. Lin and L. Susskind, {Infinite Temperature’s Not So Hot}, arXiv:2206.01083.
	\bibitem{Rahman} A.~Rahman, {dS JT Gravity and Double-Scaled SYK}, arXiv:2209.09997.
	\bibitem{Susskind4} L.~Susskind, {De Sitter Space, Double-Scaled SYK, and the Separation of Scales in the Semiclassical Limit}, arXiv:2209.09999.
	\bibitem{Narovlansky} V. Narovlansky and H. Verlinde, {Double-scaled SYK and de Sitter Holography}, arXiv:2310.16994.
	\bibitem{Verlinde} H. Verlinde and M. Zhang, {SYK Correlators from 2D Liouville-de Sitter Gravity}, arXiv:2402.02584.
	\bibitem{Jiuci1} A. Milekhin, J. Xu, {Revisiting Brownian SYK and its possible relations to de Sitter}, arXiv:2312.03623.
	\bibitem{Jiuci2} J. Xu, {Von Neumann Algebras in Double-Scaled SYK}, arXiv:2403.09021.
	\bibitem{A.Streicher} A. Streicher,  {SYK Correlators for All Energies}, J. High Energ. Phys. \textbf{2022}, 48 (2022), arXiv:1911.10171.
	\bibitem{C.Choi} C. Choi, M. Mezei and G. Sarosi,  {Exact four point function for large q SYK from Regge theory}, J. High Energ. Phys. \textbf{2021}, 166 (2021), arXiv:1912.00004.
	\bibitem{Goel} A. Goel, V. Narovlansky, and H. Verlinde, {Semiclassical geometry in double-scaled SYK}, arXiv:2301.05732.
	\bibitem{Cai} R. G. Cai, Y. H. Qi, Y. L. Wu, and Y. L. Zhang, {Topological non-Fermi liquid}, Phys. Rev. D \textbf{95}, 124026 (2017), arXiv:1601.03865.
	\bibitem{Wilczek} F. Wilczek and A. Zee, {Families from spinors}, Phys. Rev. D \textbf{25}, 553 (1982).
	\bibitem{Maldacena} J. Maldacena, G. J. Turiaci, and Z. Yang, {Two dimensional Nearly de Sitter gravity}, arXiv:1904.01911.
	\bibitem{Pethybridge} B. Pethybridge and V. Schaub, {Tensors and spinors in de Sitter space}, J. High Energ. Phys. \textbf{2022}, 123 (2022), arXiv:2111.14899.
	\bibitem{Whittaker} E. T. Whittaker and G. N. Watson, {A course of modern analysis}, Cambridge University Press, 1927.
	\bibitem{Silverman} N. N. Lebedev and R. R. Silverman, {Special functions and their applications}, Courier Corporation, 1972.
	\bibitem{Weinberg} S. Weinberg, {The Quantum theory of fields. Vol. 1: Foundations}, Cambridge University Press, 1995.
	\bibitem{Parcollet} O. Parcollet, A. Georges, G. Kotliar, and A. Sengupta, {Overscreened multichannel SU(N) Kondo model: Large-N solution and conformal field theory}, Phys. Rev. B \textbf{58}, 3794 (1998), arXiv:cond-mat/9711192.
	\bibitem{Qi} Y. H. Qi, Y. Seo, S. J. Sin, and G. Song, {Correlation functions in Schwarzian liquid}, Phys. Rev. D \textbf{99}, 066001 (2019), arXiv:1804.06164.
	\bibitem{Ahn3} Y.~Ahn, V.~Jahnke, H.~S.~Jeong, K.~Y.~Kim, K.~S.~Lee, and M.~Nishida,{Pole-skipping of scalar and vector fields in hyperbolic space: conformal blocks and holography}, J. High Energ. Phys. \textbf{2020}, 111 (2020), arXiv:2006.00974.
\end{thebibliography}
\end{document}